\documentclass[12pt,a4paper]{iopart}
\usepackage{mathptmx}
\usepackage[utf8]{inputenc}
\usepackage{color}
\usepackage{array}
\usepackage{refstyle}
\usepackage{textcomp}
\usepackage{amsbsy}
\usepackage{amstext}
\usepackage{graphicx}
\usepackage{subscript}
\usepackage{microtype}
\usepackage[unicode=true,
 bookmarks=true,bookmarksnumbered=true,bookmarksopen=false,
 breaklinks=false,pdfborder={0 0 0},pdfborderstyle={},backref=false,colorlinks=true]
 {hyperref}
\hypersetup{
 citecolor=blue, linkcolor=blue, urlcolor=blue}

\makeatletter

\AtBeginDocument{\providecommand\secref[1]{\ref{sec:#1}}}
\AtBeginDocument{\providecommand\Secref[1]{\ref{Sec:#1}}}
\AtBeginDocument{\providecommand\figref[1]{\ref{fig:#1}}}
\AtBeginDocument{\providecommand\eqref[1]{\ref{eq:#1}}}
\AtBeginDocument{\providecommand\tabref[1]{\ref{tab:#1}}}
\pdfpageheight\paperheight
\pdfpagewidth\paperwidth

\usepackage{iopams}
\usepackage{setstack}

\usepackage[all]{nowidow}
\usepackage{multicol}
\usepackage{multirow}
\usepackage{booktabs}
\usepackage[font=small,labelfont=bf]{caption}
\AtBeginDocument{%
\renewbibmacro{in:}{\ifentrytype{article}{}{\printtext{\bibstring{in}\intitlepunct}}}
\DeclareFieldFormat[article]{title}{#1}
\DeclareFieldFormat[inproceedings]{title}{#1}
}

\@ifundefined{showcaptionsetup}{}{%
 \PassOptionsToPackage{caption=false}{subfig}}
\usepackage{subfig}
\makeatother

\usepackage[bibstyle=numeric,citestyle=authoryear,maxcitenames=1,uniquelist=false,firstinits=true,backend=biber,giveninits=true,doi=false,url=true,isbn=false]{biblatex}
\begin{document}

\title[Clinical prompt gamma-ray spectroscopy for proton range verification]{A full-scale clinical prototype for proton range verification using
prompt gamma-ray spectroscopy}

\author{Fernando~Hueso-González$^{1}$, Moritz~Rabe$^{1}$, Thomas A Ruggieri$^{1}$,
Thomas~Bortfeld$^{1}$, Joost M Verburg$^{1}$ }

\address{$^{1}$Massachusetts General Hospital and Harvard Medical School,
Department of Radiation Oncology, Boston, MA 02114, USA}

\ead{jverburg@mgh.harvard.edu}
\begin{abstract}
We present a full-scale clinical prototype system for in vivo range
verification of proton pencil-beams using the prompt gamma-ray spectroscopy
method. The detection system consists of eight LaBr\textsubscript{3}
scintillators and a tungsten collimator, mounted on a rotating frame.
Custom electronics and calibration algorithms have been developed
for the measurement of energy- and time-resolved gamma-ray spectra
during proton irradiation at a clinical dose rate. Using experimentally
determined nuclear reaction cross sections and a GPU-accelerated Monte
Carlo simulation, a detailed model of the expected gamma-ray emissions
is created for each individual pencil-beam. The absolute range of
the proton pencil-beams is determined by minimizing the discrepancy
between the measurement and this model, leaving the absolute range
of the beam and the elemental concentrations of the irradiated matter
as free parameters. The system was characterized in a clinical-like
situation by irradiating different phantoms with a scanning pencil-beam.
A dose of 0.9 Gy was delivered to a $5\times10\times10$ cm\textsuperscript{3}
target with a beam current of 2 nA incident on the phantom. Different
range shifters and materials were used to test the robustness of the
verification method and to calculate the accuracy of the detected
range. The absolute proton range was determined for each spot of the
distal energy layer with a mean statistical precision of 1.1 mm at
a 95\% confidence level and a mean systematic deviation of 0.5 mm,
when aggregating pencil-beam spots within a cylindrical region of
10 mm radius and 10 mm depth. Small range errors that we introduced
were successfully detected and even large differences in the elemental
composition do not affect the range verification accuracy. These results
show that our system is suitable for range verification during patient
treatments in our upcoming clinical study.
\end{abstract}
\maketitle

\section{Introduction}

The advantageous depth-dose deposition profile of protons cannot yet
be fully utilized in clinical practice because of uncertainty in the
beam range within the patient. The origin of this uncertainty is the
determination of the stopping power of tissue along the beam path,
which is affected by the CT imaging, the degeneracy in the conversion
from x-ray attenuation to stopping power, and the uncertainty in the
ionization potential of human tissue \parencite{andreo_range}. Furthermore,
interfractional and intrafractional patient positioning errors and
anatomical changes can affect the range. Conservative safety margins
\parencite{albertini_margins} and robust treatment planning \parencites{unkelbach_robust}
are currently required to ensure a complete tumor coverage.

Efforts are being undertaken to reduce proton range uncertainty and
the required safety margins by performing an in vivo verification
of the range during patient treatment. Several methods have been suggested
\parencite{range_review}, but no solution is available for routine
clinical use. The use of prompt gamma-rays, which originate from proton-nuclear
interactions with tissue, was proposed by \textcite{jongen_pg} and
is being actively investigated by various groups \parencite{krimmer_review}.
The virtually instantaneous emission of the gamma-rays, with a time
scale smaller than $10^{-11}\,\mathrm{s}$, enables real-time range
verification of the delivered proton pencil-beams. Through the high
gamma-ray energy of typically several MeV, most of them escape the
patient and can be detected externally. For a successful clinical
application, a detection system has to meet several requirements.
It needs to detect a sufficient number of gamma-rays during a fractionated
clinical treatment with fields that deliver a dose on the order of
$1\,\mathrm{Gy}$ to the tumor, such that the range can be determined
with millimeter precision. The correlation of the measured gamma-ray
emissions with the proton dose deposition needs to be accurately known
to minimize systematic bias. Determining this correlation based on
first principles requires extensive modeling of the nuclear reactions
\parencite{verburg_spec}, but more empirical methods are also being
investigated \parencite{schumann_pgdose}. To not extend the duration
of the treatments, the detection needs to be performed while the protons
are delivered at a standard clinical beam current of about 2 nA incident
on the patient. Under these conditions and with a sufficiently efficient
system, a detector load of millions of gamma-rays per second is expected.
The detection system also needs to function stably under the highly
variable count rates that occur during proton pencil-beam delivery.
These are certainly challenging requirements.

Individual prompt gamma-rays can be detected with scintillation detectors
\parencites{first_pgexp}{smeets_slit}{pinto_multislit}{krimmer_pgpi}{pausch_scint_u100_tns}{verburg_pg},
which need to be dense and thick to absorb the full energy. A form
of collimation is needed to reconstruct the spatial origin of the
emission. Collimation is only needed along the depth dimension, because
the lateral position of each incident pencil-beam is monitored with
an ionization chamber in the beam line. A physical collimator made
of a high-Z material, with one or multiple slits perpendicular to
the proton beam direction, limits the longitudinal position along
the beam path from where the detected gamma-rays originate. Another
advantage of a physical collimator is the shielding of gamma-rays
emitted in the beam entrance path, which do not provide valuable information
about the proton range. Alternatively, the transit time of the protons
in tissue until the prompt gamma-ray is emitted can be used as a form
of virtual collimation \parencite{golnik_pgt,hueso_pgt}. Such time-based
collimation reduces weight and adds geometric flexibility. However,
because of the time spread of the incident protons in a bunch from
a clinical proton acceleration system \parencite{petzoldt_pbm}, the
origin of the gamma-ray emission is less well defined than for passive
collimation. It also adds uncertainty because of the need to correlate
the proton transit time with a spatial location in the patient. Finally,
the application of Compton cameras is being investigated \parencite{hueso_bbcc,solevi_macaco,aldawood_cc,rohling_simu_cc,draeger_polaris}.
These systems do not require physical collimation either, but rely
on the detection of multiple Compton scatter interactions per incident
gamma-ray in order to reconstruct its origin region.

Prompt gamma-ray imaging is an indirect range verification method
that exploits the correlation of the nuclear reaction positions with
the deposited dose. Nuclear interactions between accelerated protons
and the nuclei of human tissue are complex processes. Numerous nuclear
reaction channels of protons with the main nuclei of tissue, $\mathrm{^{12}C}$
and $\mathrm{^{16}O}$, exist. The residual nuclei de-excite through
the emission of gamma-rays with different discrete energies. The patient-specific
emissions of prompt gamma-rays during the treatment can be modeled
based on the CT scan, the treatment plan and nuclear physics models.
Matching this model with the measured prompt gamma-ray distribution
provides a means to verify the ranges of the proton pencil-beams. 

In this context, it is important to define the goal of the range verification
method. One approach is to detect a deviation in the beam range through
the comparison of measurements from different treatment days. Such
a detection would be clinically useful for quality assurance, to verify
the consistency among the delivered fractions. However, this approach
does not provide an estimate of the absolute proton range. Reducing
the initial range margins, which is our goal, imposes a more difficult
requirement on the range verification. It requires an accurate and
robust determination of the absolute range of the beam with only the
treatment plan as prior knowledge.

Recently, initial studies of prompt gamma-ray detection during patient
treatments with passively scattered protons \parencite{richter_slit}
and actively scanned proton pencil-beams \parencite{xie_slit} were
performed with a scintillation detector array combined with a knife-edge
slit collimator. \textcite{xie_slit} measured the integral number
of prompt gamma-rays during the pencil-beam delivery. The median statistical
precision of the range verification for a set of distal layers was
$\pm$12 mm at 95\% confidence ($2\sigma$) for individual pencil-beams,
and $\pm$4.2 mm after aggregating pencil-beams laterally within the
same layer with a Gaussian kernel with a 7 mm standard deviation.
The positioning accuracy of the setup was estimated as $\pm$1.5 mm
at $2\sigma$ \parencite{xie_slit}. The bias in the absolute range
from sources other than positioning was not discussed. In the case
of Compton cameras, relative range shifts of a few millimeters are
theorized to be detectable with larger scale Compton camera systems
\parencite{draeger_polaris}, which have not been constructed yet.

For several years, we have been developing a proton range verification
method that is uniquely based on prompt gamma-ray spectroscopy \parencite{verburg_pg,verburg_simu,verburg_spec}.
This method uses both the arrival time and energy of the detected
prompt gamma-rays. Using a small-scale prototype, we first showed
energy- and time-resolved prompt gamma-ray spectra that were measured
during the delivery of proton beams \parencite{verburg_pg}. We use
detailed models of the nuclear reactions associated with the discrete
prompt gamma-ray emissions to verify the absolute range of proton
beams in phantoms, without requiring prior knowledge of the elemental
composition of the irradiated matter \parencite{verburg_spec}. Moreover,
the measurement of the arrival time of the gamma-rays allows a separation
of proton- and neutron-induced gamma-rays, which removes the confounding
uncertainty from the neutron-induced gamma radiation background in
the treatment room.

In this paper, we describe the development and evaluation of a new
full-scale clinical prototype detection system for proton range verification
based on prompt gamma-ray spectroscopy. This system will be deployed
for the first clinical study with patients. We present the detectors,
collimator and data processing methods in \secref{DetectorsData}.
\Secref{RangeVerification} describes a fully integrated clinical
workflow for modeling the prompt gamma-ray emissions based on the
treatment plan, in order to predict the number of gamma-rays detected
when different range errors occur, and to verify the proton range
by comparing the measurements with the model. The experimental setups
that we use to optimize the cross section data and to evaluate the
range verification performance of the system are shown in \secref{ExperimentalSetups}.
The results of the experiments are given in \secref{Results}, in
which we assess the detector performance and the accuracy and statistical
precision of the proton range verification. In \secref{Discussion},
we discuss the obtained results and the consequences for the clinical
implementation, and the main conclusions of the study are drawn in
\secref{Conclusions}.

\section{Detectors and data processing\label{sec:DetectorsData}}

\subsection{Proton beam}

The experiments were performed in the pencil-beam scanning treatment
gantry at the Francis H. Burr Proton Therapy Center, Massachusetts
General Hospital. Protons were accelerated to 230 MeV with an IBA
C230 cyclotron (Ion Beam Applications SA, Louvain-la-Neuve, Belgium),
degraded by an energy selection system, and transported to the treatment
room along a beam line with a length of approximately 35 meters. The
cyclotron features a radiofrequency of 106.3 MHz. All pencil-beam
layers were delivered with the standard clinical system at the full
beam current of 2 nA incident on the phantom, which corresponds to
a bunch of approximately 100 protons every 9.4 ns. Based on routine
quality assurance measurements, the reproducibility of the beam range
in water is known to be better than 0.5 mm.

\subsection{Detectors and collimator}

Gamma-rays are detected using eight detector modules, each consisting
of a cerium-doped lanthanum(III) bromide scintillation crystal (Saint-Gobain,
Saint Pierre Lès Nemours, France) with a diameter of 50.8 mm and a
length of 76.2 mm, a photomultiplier tube, and a custom base with
a Cockcroft–Walton high voltage generator and a transistorized voltage
divider. The total volume of scintillation material is 1236 cm\textsuperscript{3}.
The detector signals from the anode of the photomultipliers are amplified
with transimpedance amplifiers that are located next to the detectors.
The scintillators and electronics are designed to sustain high overall
event rates of up to $10^{7}$ events per second.

\begin{figure}
\begin{centering}
\includegraphics[height=6cm]{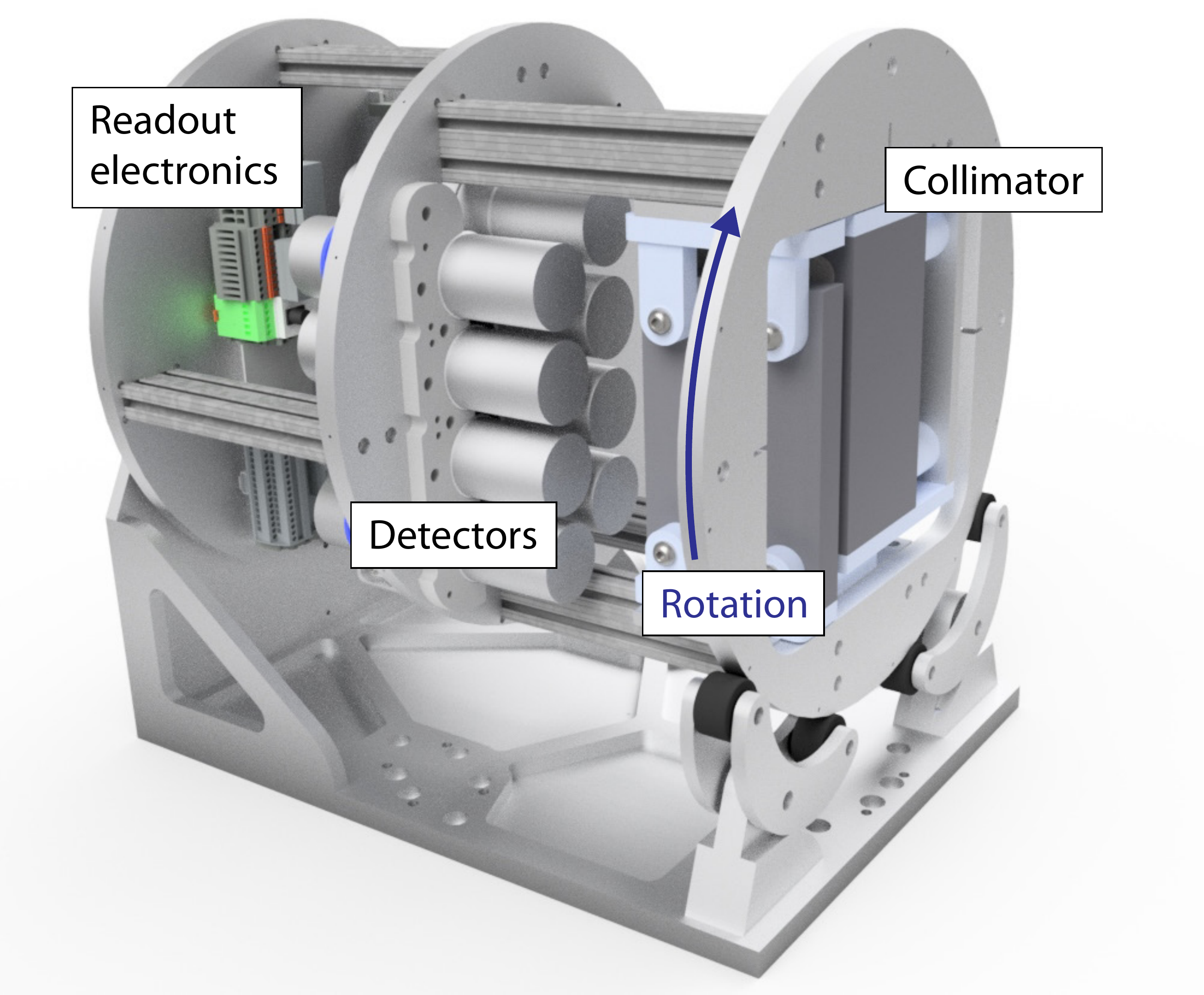}\hfill{}\includegraphics[height=6cm]{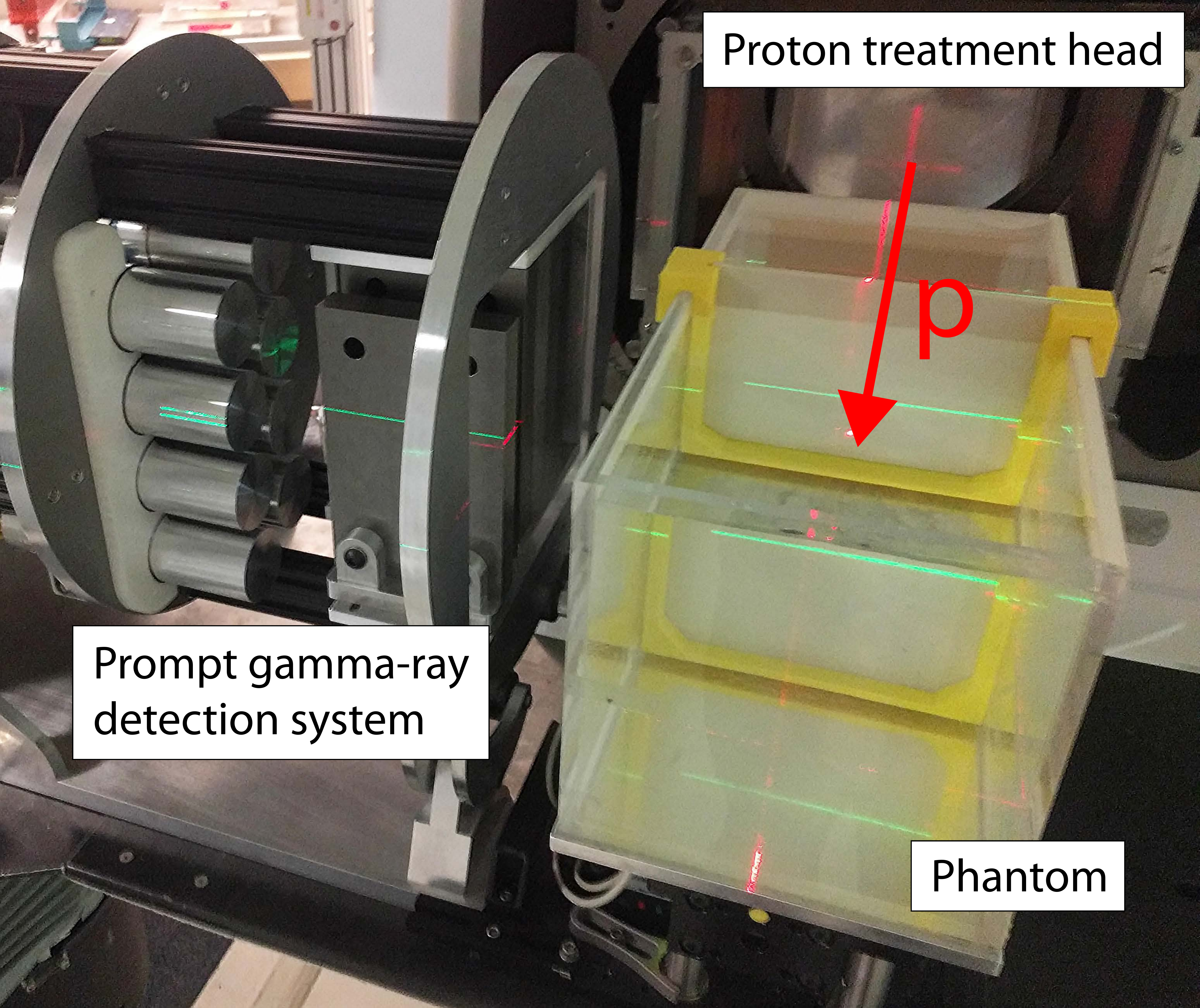}
\par\end{centering}
\textsf{\caption{Left: 3D model of the clinical prototype system, which can rotate
around its axis according to the beam incidence angle. The tungsten
collimator is visible on the front plane, the eight scintillation
detectors in the middle and the readout electronics on the back plate.
Right: photo of the system in the gantry treatment room. The red arrow
shows the proton beam incidence direction. }
\label{fig:detector} }
\end{figure}

The gamma-rays in the entrance path of the beam are shielded by a
127 mm thick and 102 mm wide block of tungsten. Along the beam direction,
this is followed by a slit opening of 12.7 mm and a single collimator
slab with a width of 25.4 mm. Four detectors are stacked with their
center aligned with the edge of the proximal collimator. The other
four detector modules are located distal of these detectors in a closely
packed configuration (see \figref{detector}). Each row of detectors
therefore focuses on a different position along the beam direction.
Because many of the prompt gamma-rays interact through Compton scattering
or pair production, the volume of scintillation material that is not
directly in view of the primary prompt gamma-rays contributes to the
probability of the absorption of the full photon energy.

\subsection{Data acquisition system}

The detectors are connected to a custom data acquisition system located
in the treatment control room. This system is synchronized with the
pencil-beam delivery system and the cyclotron. The signals from the
eight detector modules are independently read out with a 14-bit analog-to-digital
converter that is phase-locked to the accelerator at a frequency of
twice the cyclotron radiofrequency, resulting in a sample rate of
approximately $2.13\times10^{8}$ samples per second. The data from
the analog-to-digital converters are processed in real-time by field-programmable-gate
arrays (FPGAs). A trigger to acquire an event is generated when the
digitized signal magnitude exceeds a predefined threshold.

The gamma-ray energy is calculated by integrating the area under the
pulse signal over a time window of 200 ns. The pulse shape is used
to determine whether the gamma-ray event is a single gamma-ray detection
or a pile-up of several gamma-rays (\secref{PileUpRejection}). The
calculations and the storage in memory result in a dead-time of 150
ns after each event, during which the individual detector module is
not sensitive to a new event. Because of the fast detectors and the
short dead-time, the average separation time between two gamma-ray
events is much longer than the time required to acquire an event.
We can typically acquire about 90\% of the gamma-rays that interact
with the detectors and deposit an energy within the range of interest.

To find the precise arrival time of the gamma-ray, we subtract a delayed
copy of the digital signal from the original one and find the zero
crossing time of the resulting signal. This time point is independent
of the signal magnitude. The arrival time analysis is first performed
with a time resolution equal to the sample rate. Subsequently, we
use a polynomial interpolation to achieve sub-nanosecond resolution.

A second data acquisition board performs real-time acquisition of
the pencil-beam delivery status and the signals from the beam monitoring
electronics. It also enables the detector readout only during the
beam-on periods. The clock of this board is synchronized with the
detector readout board, therefore providing the exact beam status
of the pencil-beam delivery for each detected gamma-ray.

The acquired data are stored in a memory buffer that is continuously
read out by the control software, which is a custom C++ application
on a Linux system. This application provides a real-time graphical
display of the acquired data and the beam status. An important task
of this software is to perform various corrections to normalize all
measurements to absolute units, independently of conditions such as
the detector load and the neutron-induced background, as will be discussed
in the next sections.

\subsection{Energy calibration}

Although the detector readout electronics are designed for high stability
at various detector count rates, a small dependency of the signal
amplitude as a function of the count rate is unavoidable. Moreover,
any scintillator has some temperature dependence. To make the measurements
independent of these factors, an automated energy calibration as a
function of the detector count rate is implemented. After applying
the initial fixed calibration factor, an energy correction function
is determined for each proton energy layer using the gamma-rays from
neutron capture by hydrogen at 2.22 MeV, which are very abundant in
every measurement. A 2D histogram with the uncorrected line energy
as a function of the pencil-beams ordered by count rate is created.
A 2D curve is fitted, namely a normal distribution in the energy dimension
with a shifting mean value, which is linearly correlated to the ordered
pencil-beam number. The benefit of simultaneously considering all
pencil-beams within a layer is that statistical variations have negligible
impact. The ratio between the uncorrected energy at the peak position
and the actual gamma-ray energy of the neutron-hydrogen line is applied
as a small correction on the order of 1\%.

A very small non-linearity is also present in the correlation of the
signal amplitude with energy. The correction for this effect is also
automated and is the same for all pencil-beams. This correction is
performed by analyzing the spectral position of a few strong gamma-ray
lines with known energies and fitting a polynomial correction function,
which is then applied to all measurements.

\subsection{Time calibration}

The gamma-ray arrival time relative to the cyclotron radiofrequency
period is different for each energy layer because of the different
flight times of the protons through the beam lines. For each detector
and spot, we automatically center the proton-induced prompt gamma-ray
peak in the time dimension. Due to space charge effects in the photomultiplier,
the shape of the detector signals slightly depends on the energy.
To remove any energy dependency in the arrival time, a correction
is applied by finding the location of the proton-induced peak in the
time dimension for several energy regions and by performing a linear
fit. This is similar to the method described by \textcite{hueso_elbe}.

\subsection{Pile-up rejection\label{sec:PileUpRejection}}

Piled-up events, where multiple gamma-rays interact with the detector
at almost the same time such that they can not be separated, are detected
using pulse shape analysis. A set of expected pulse shape parameters
and their variance were obtained from a laboratory measurement at
a very low event rate. Whenever one parameter is outside the expected
nominal window, i.e., the regular pulse shape, the event is marked
as a piled-up event and discarded from further analysis. The acceptance
window is calibrated for each detector separately to account for the
slight differences in pulse shape between detectors.

A small fraction of the piled-up events cannot be detected by this
method. This can happen if two events are very close to each other
in time $(T_{1}=15\,\text{{ns}}$) so that the sum of pulse shapes
is similar to a single one, independently of their respective energies;
or if the second event arrives within $T_{2}=150\,\text{{ns}}$, but
has a small energy with respect to the previous one, so that the pulse
shape difference is not significant. Consequently, events with a wrong
energy or time stamp are included in the analysis as valid events,
but contribute to the background rather than to resolved lines. A
correction scale factor $S$ is applied depending on the detector
count rate $R$:
\begin{equation}
S=\exp[R(f_{1}\times T_{1}+f_{2}\times T_{2})],\label{eq:fnc}
\end{equation}
in which $f_{1}$ and $f_{2}$ are empirically determined parameters
for the two aforementioned effects and $T_{1}$ and $T_{2}$ are the
corresponding time constants. The values of these parameters are determined
such that the magnitude of the proton-induced gamma-rays acquired
in a fixed time period is proportional to the proton beam current.

\subsection{True number of events\label{sec:TriggerRate}}

The number of valid events $N_{\mathrm{{E}}}$ during the delivery
of a pencil-beam is lower than the true number of events $N_{\mathrm{{T}}}$
because of the dead-time and pile-up of two or more gamma-rays. To
determine the true number of events, we take advantage of the fact
that Poisson statistics apply to the occurrence of the events. This
assumption is valid if the dose rate is constant, which is a good
approximation because the duration of the pencil-beam delivery is
much longer than the duration of the ramp-up and ramp-down processes.
As described by \textcite{becares_truerate}, Poisson statistics imply
that the separation times between consecutive events feature an exponential
distribution. We fit the measured separation time distribution for
times that exceed the pulse duration and dead-time. The true number
of events is then given by the integral of the fitted distribution,
including also the virtual events with a short separation time that
precluded them from being acquired. The correction is applied as a
scaling factor $D=N_{\mathrm{{T}}}/N_{\mathrm{{E}}}$ equal to the
ratio between true and acquired events.

\subsection{Coincidence rejection}

When analyzing discrete gamma-ray lines, it is desirable to reject
events in which Compton scatter occurred in the scintillator. In these
events, the full energy of the gamma-ray was not absorbed in a single
detector and therefore the event contributes only to the continuum
background. If the scattered gamma-ray interacts with a second detector,
we are able to reject such events through the analysis of coincidences
between the detectors. If events in separate detectors are separated
in time by less than 3 ns, both events are rejected.

There is a small probability of erroneously rejecting events that
are not due to Compton scatter. This is corrected for each pencil-beam
by using a random coincidence correction, based on a histogram of
the time difference between consecutive events in different detectors.
It is expected to present a baseline plus a peak of true coincidences
around zero. The total number of coincidences $N_{\mathrm{{C}}}$
is the integral of the histogram within the coincidence window. The
number of random coincidences $N_{\mathrm{{R}}}$ is determined by
fitting the baseline of the histogram with a quadratic curve and calculating
its integral within the aforementioned window. The random coincidence
correction $C$ is calculated as:

\begin{equation}
C=1+\frac{1/N_{\mathrm{{E}}}}{1/N_{\mathrm{{R}}}-1/N_{\mathrm{{C}}}}.\label{eq:rcc}
\end{equation}

\subsection{Histogram analysis\label{sec:HistogramAnalysis}}

The fully corrected data are combined in 2D histograms with dimensions
of energy and time \parencite{verburg_pg}. We create two histograms
for each proton pencil-beam: one for each row of four detectors, see
\figref{detector}. An iterative method is used to split the measured
2D histograms into three different components:
\begin{description}
\item [{Proton-induced~continuum.}] Prompt gamma-rays that undergo incoherent
scattering in the detector, or that have scattered before reaching
the detector, contribute to an unresolved background in the energy
dimension, but are well resolved in the time dimension. A second contribution
to the unresolved continuum is from a quasi-continuum of discrete
gamma-rays from the cascade decay of higher excited states \parencite{ripl},
which cannot be resolved as separate lines.
\item [{Neutron-induced~continuum.}] Like the proton-induced continuum,
the neutron-induced gamma-rays can also scatter before or inside the
detector, or stem from unresolved cascades. This structure is both
unresolved in time and energy.
\item [{Resolved~lines.}] Subtracting the continuum components results
in a histogram with only resolved lines. They correspond to prompt
gamma-rays that either deposit the full energy in one of the detectors,
or the full energy minus one or two 511 keV escape photons. These
peaks are resolved in both the energy and time dimension. In addition,
neutron-induced reactions such as the neutron capture of hydrogen
result in lines that are resolved in energy but not strongly resolved
in time.
\end{description}
A robust separation of these three components is crucial, because
it allows a direct comparison of the measurements with fundamental
physics models. The discrete proton-induced gamma-ray lines can be
fully modeled based on nuclear reaction cross sections, and their
quantification also carries the essential information about the elemental
concentrations in the irradiated tissue. For a spot-wise 2D separation
of proton- and neutron-induced events, a dedicated algorithm was designed:
the Recursive Bisection Neutron Subtraction (ReBiNS). First, a time
projection of the 2D histogram is obtained for all energy bins. The
neutron background is estimated on the resulting 1D time spectrum
$M(t)$ by means of the SNIP algorithm \parencite{ryan_snip}. The
proton-induced histogram is obtained by subtracting the neutron background
from the original histogram $M(t)$. Both the neutron-induced and
proton-induced 1D histograms are normalized and fitted by cubic splines,
serving as numeric probability density functions of the neutron $s_{n}(t)$
and proton $s_{p}(t)$ time structures. Then, the time spectrum $M(t)$
is fitted to
\begin{eqnarray}
M(t) & = & N\times s_{n}(t)+P\times s_{p}(t),\label{eq:timemodel}
\end{eqnarray}

where $N$ is the number of neutron-induced counts, $P$ the number
of proton-induced counts, and $N+P$ is constrained to the total number
of histogram entries $T$ in $M$. Once the parameters are calculated,
a similar fit is done for each energy bin individually. To ensure
the robustness and consistency with the global fit, a recursive bisection
strategy is applied. First, two time projections $a$ and $b$ are
created among the upper and lower half of the 2D energy-over-time
spectrum. Then, a simultaneous fit of \eqref{timemodel} is done on
both projections, but with fixed shapes $s_{n}(t)$, $s_{p}(t)$ and
the cross-constraints $N_{a}+N_{b}=N$, $P_{a}+P_{b}=P$, $N_{a}+P_{a}=T_{a}$,
$N_{b}+P_{b}=T_{b}$. Each of the regions is again divided in two
sub-regions, and the procedure is repeated recursively until the sub-regions
consist of single energy bins $i$. The set of fitted parameters $N_{i}$
and $P_{i}$ can be interpreted as an energy spectrum of the neutron-induced
and proton-induced contributions. Both 1D spectra are further analyzed
to separate the continuum background from the resolved lines by means
of the SNIP algorithm. 

The magnitude of the gamma-ray lines is determined by a 2D fit of
the resolved line histogram. The peaks are modeled as Gaussians in
the energy dimension and as $s_{p}(t)$ in the time dimension, or
$s_{n}(t)$ in the case of resolved neutron-induced lines.

\subsection{Lateral spot merging\label{sec:LateralMerging}}

For range verification, we create histograms in which the data of
neighboring pencil-beams within the same energy layer are summed if
their center position is within a 10 mm lateral radius of the spot
under consideration. With our beam delivery system, this usually results
in data from seven pencil-beams to be accumulated. Based on Poisson
statistics, one would theoretically expect this to yield an improvement
of the statistical precision by a factor of 2.6, at the expense of
a reduction in spatial resolution. Note that considerable overlap
already exists between neighboring pencil-beams near the end-of-range
because of multiple Coulomb scattering.

\section{Range verification method \label{sec:RangeVerification}}

A sophisticated simulation of the fundamental physical processes like
nuclear interactions, attenuation and detector system response was
developed to accurately predict the prompt gamma-ray emissions and
detections. This model is compared to the measurement data to determine
the proton range for each pencil-beam.

\begin{figure}
\begin{centering}
\includegraphics[scale=0.75]{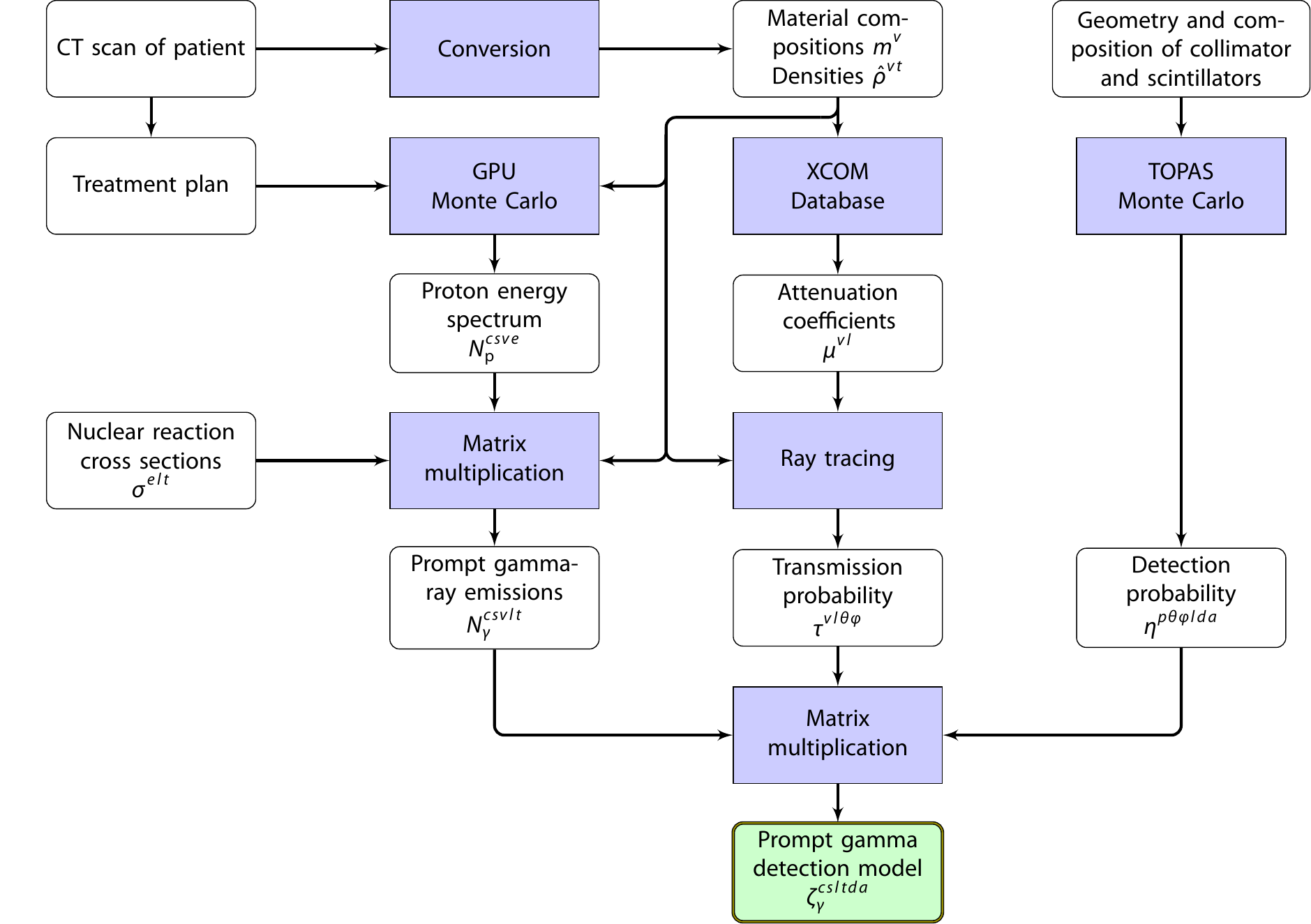}
\par\end{centering}
\caption{Simplified workflow chart visualizing the prompt gamma-ray detection
model generation based on the CT scan of the patient, the treatment
plan, the detector geometry, the tabulated nuclear cross sections
and the XCOM database. GPU and TOPAS Monte Carlo simulations, as well
as analytical ray tracing are performed. The mathematical symbols
are described in \tabref{overviewParameters}.}

\label{fig:flowchart}
\end{figure}

The model is split up into several parts: geometry definition, CT
to material conversion, proton stopping process, prompt gamma-ray
emission, gamma-ray attenuation, and detection. An overview of the
indices and parameters defined in subsequent sections is presented
in \tabref[s]{overviewIndices} and \ref{tab:overviewParameters},
and a graphical workflow is shown in \figref{flowchart}.

\begin{table}
\caption{Indices used within the prompt gamma-ray emission and detection model.\label{tab:overviewIndices}}

{\small{}\small{}%
\begin{tabular}{>{\raggedright}p{0.05\columnwidth}>{\raggedright}p{0.08\columnwidth}>{\raggedright}p{0.08\columnwidth}>{\raggedright}p{0.09\columnwidth}>{\raggedright}p{0.1\columnwidth}>{\raggedright}p{0.45\columnwidth}}
\hline 
\textbf{Index } & \textbf{Bins } & \textbf{Step } & \textbf{Min.} & \textbf{Max. } & \textbf{Description }\tabularnewline
\hline 
$v$ & \# voxels & 1 & 0 & \# voxels  & voxel of the field of view ($120\times160\times160$)\tabularnewline
$p$ & \# pixels  & 1 & 0 & \# pixels & pixel of the front plane ($120\times160$)\tabularnewline
$e$  & 150  & 1 MeV  & 0 MeV  & 150 MeV & proton kinetic energy within the field of view\tabularnewline
$s$  & \# spots & 1  & 0 & \# spots & spot of the treatment plan\tabularnewline
$l$  & 7  & 1  & 0  & 6  & prompt gamma-ray line (0: 1.6 MeV, 1: 2.0 MeV, 2: 2.3 MeV, 3: 2.8
MeV, 4: 4.4 MeV, 5: 5.2 MeV, 6: 6.1 MeV)\tabularnewline
$t$  & 2  & 1  & 0  & 1  & target nucleus (0: oxygen $^{16}$O, 1: carbon $^{12}$C)\tabularnewline
$\theta$  & 360  & 0.25° & 0° & 90° & polar angle\tabularnewline
$\phi$  & 360  & 1.00° & 0° & 360° & azimuthal angle\tabularnewline
$a$  & 5  & 1  & 0  & 4  & absorption process: (0) full, (1) single and (2) double escape, (3):
any energy deposit between 1.4 MeV and 3.0 MeV, or (4): higher than
3.0 MeV.\tabularnewline
$d$  & 8  & 1  & 0  & 7  & scintillation detector\tabularnewline
$r$  & 2  & 1  & 0  & 1  & row of detectors ($0:\sum_{d=0}^{3}$, $1:\sum_{d=4}^{7}$)\tabularnewline
$c$  & 17  & 1  & 0  & 16 & discrete range error scenario in steps of 1 mm ($c=8\rightarrow$nominal
scenario)\tabularnewline
\hline 
\end{tabular}{\small{}}}{\small\par}
\end{table}

\begin{table}
\caption{Parameters used within the prompt gamma-ray emission and detection
model.\label{tab:overviewParameters}}

{\small{}\small{}%
\begin{tabular}{>{\raggedright}p{0.12\columnwidth}>{\raggedright}p{0.58\columnwidth}>{\raggedright}p{0.2\columnwidth}}
\hline 
\textbf{Parameter} & \textbf{\small{}Description } & \textbf{\small{}Source }\tabularnewline
\hline 
$m^{v}$  & material composition in voxel $v$. & CT, \textcite{Schneider00} \tabularnewline
$\mu^{vl}$  & attenuation coefficient in voxel $v$ in cm\textsuperscript{-1} for
a $\gamma$-ray with energy $l$. & $m$, XCOM \cite{nist_xcom}\tabularnewline
$\rho^{vt}$  & concentration by mass of target $t$ in voxel $v$ in g/cm\textsuperscript{3}.
$\hat{\rho}^{vt}$: expected, $\rho^{vt}$: optimized.  & CT \tabularnewline
$N_{\mathrm{p}}^{csve}$  & protons in voxel $v$ with energy $e$ for spot $s$ and range error
scenario $c$. & GPU\tabularnewline
$\sigma^{elt}$  & differential cross section at 90 degrees of a proton with energy $e$
reacting with target $t$ and emitting a prompt $\gamma$-ray with
energy $l$. $\hat{\sigma}^{elt}$: expected, $\sigma^{elt}$: optimized. & \textcite{verburg_spec} \tabularnewline
$\psi^{eta}$  & effective cross section accounting for all prompt $\gamma$-rays from
unresolved energy levels $l>6$ at target $t$ for absorption processes
$a=3,4$ and proton energy $e$. & Experiment \tabularnewline
$N_{\gamma}^{csvlt}$  & prompt $\gamma$-rays of transition $l$ of nucleus $t$ emitted in
voxel $v$ for spot $s$ and range error scenario $c$. & $N_{\mathrm{p}}\rho\sigma$ \tabularnewline
$\tau^{vl\theta\phi}$  & probability that a prompt $\gamma$-ray with energy $l$ emitted in
voxel $v$ in direction $(\theta,\phi)$ reaches the front plane $z=160$
mm without interaction. & $\mu$, ray tracing\tabularnewline
$\eta^{p\theta\phi lda}$  & probability that a prompt $\gamma$-ray with energy $l$ crossing
the pixel $p$ of front plane $z=160$ mm at an angle $(\theta,\phi)$
undergoes absorption process $a$ in detector $d$. & TOPAS \cite{TOPAS12}\tabularnewline
$\xi^{vlda}$  & probability that a prompt $\gamma$-ray with energy $l$ emitted in
voxel $v$ reaches the front plane without undergoing any interaction
and then undergoes absorption process $a$ in detector $d$. & $\eta\tau$\tabularnewline
$\zeta_{\gamma}^{csltda}$  & expected number of prompt $\gamma$-rays undergoing absorption process
$a$ in detector $d$, that were emitted from target nucleus $t$
and from transition $l$ for spot $s$ and range error scenario $c$. & $N_{\gamma}\xi$\tabularnewline
$C_{\gamma}^{slra}$  & measured $\gamma$-rays from transition $l$, absorption process $a$
for spot $s$ and detector row $r$. & Experiment\tabularnewline
$k^{st}$ & elemental concentration correction factor ($\rho=k\hat{\rho}$) for
target $t$ and spot $s$. & Experiment, Fit\tabularnewline
$\epsilon{}^{s}$  & absolute range error for spot $s$, namely the difference between
the optimized and the planned range $\epsilon=(c-8)\times(1\,\text{{mm})}$. & Experiment, Fit\tabularnewline
\hline 
\end{tabular}{\small{}}}{\small\par}
\end{table}

\subsection{Field of view}

\begin{figure}
\subfloat[Beam perspective\label{fig:fovBeamPerspective}]{\includegraphics[height=0.185\paperheight]{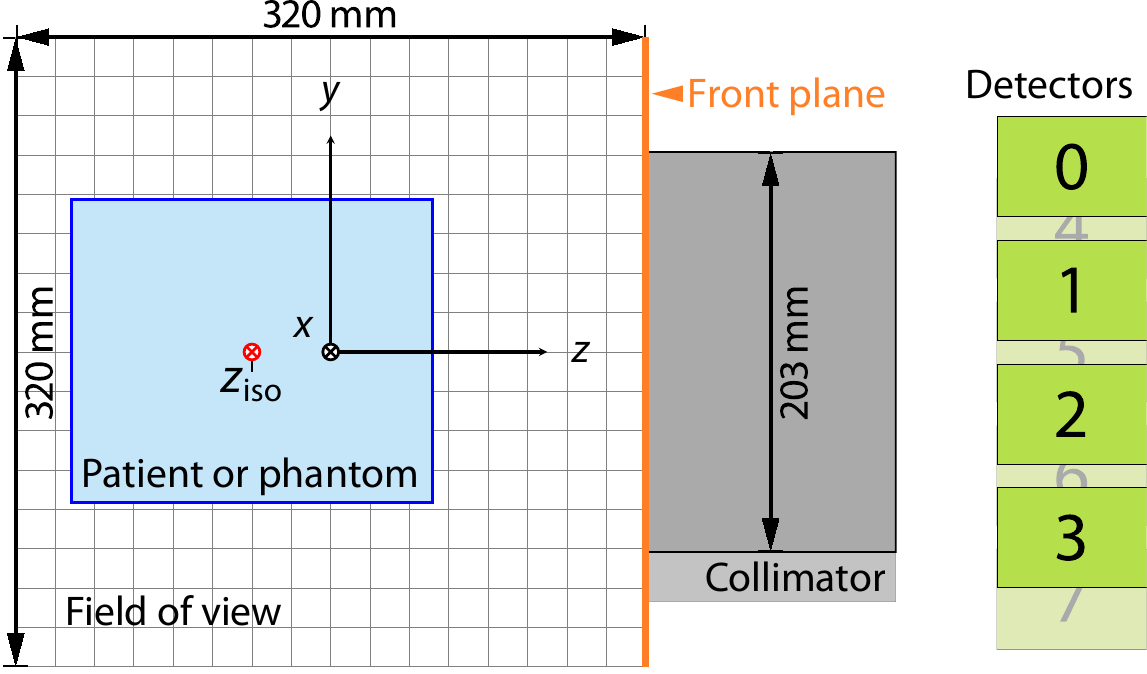}

}\hfill{}\subfloat[Detector perspective\label{fig:fovDetectorPerspective}]{\includegraphics[height=0.185\paperheight]{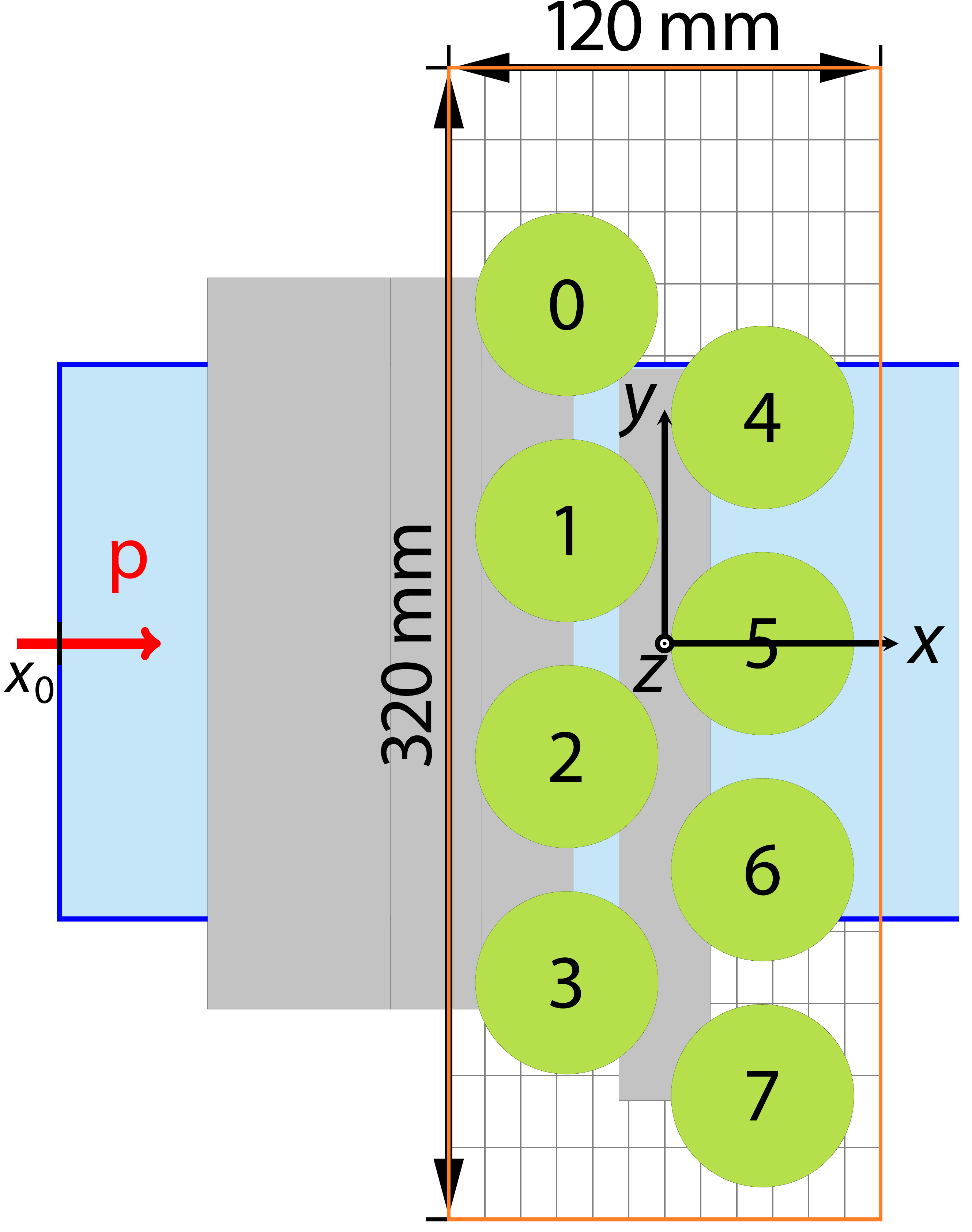}

}

\caption{Sketch of the detector FOV from the perspective of the beam (a) and
of an observer behind the detector (b). The collimator slabs (gray)
and the eight detectors (green) are overlaid. A patient or phantom
(blue) is shown as an example, that could be located anywhere within
the FOV. The $x$ axis is parallel to the beam axis (red), the $y$
axis is parallel to the slit, and the $z$ axis completes the right-handed
triad, pointing towards the detectors. The $x$ axis origin is on
the center of the standalone collimator slab at a distance $\left|x_{0}\right|$
from the upstream edge of the phantom or patient, whereas the $y$
axis is centered on the proximal collimator block. The $z$ axis origin
is at 160 mm normal to the front plane (orange line) of the collimator
and at a lateral distance $\left|z_{\mathrm{iso}}\right|$ of the
central beam axis. The FOV covers a region of $120\times320\times320\,\text{mm}^{3}$.
\label{fig:fov}}
\end{figure}

A volume of 12\,cm in the direction parallel to the proton beam path
(x axis), 32\,cm in the direction perpendicular to the pencil-beams
pointing towards the detector system (z axis) and 32\,cm in the third
dimension (y axis) is defined as the field of view (FOV) of the detector,
see \figref{fov}. This volume is discretized into voxels of $1\times2\times2\,\text{mm}^{3}$
to define a grid on which the simulation is performed. The voxel size
in the dimension parallel to the beam path is set to the smallest
distance of 1\,mm, since this axis corresponds to the direction along
which the range verification is performed. The FOV therefore consists
of $120\times160\times160$ voxels. We define the front plane of the
FOV at the surface $z=160$ mm, which is divided in $120\times160$
pixels, as seen in \figref{fov}.

\subsection{CT scan and material conversion}

As the first step of the model simulation, the CT numbers of the patient
or phantom scan are interpolated onto the FOV grid. Based on standard
conversion schemes, the following data are estimated for each voxel
$v$: the material composition $m^{v}$, the mass concentration $\hat{\rho}^{vt}$
of oxygen and carbon, and the attenuation coefficient $\mu^{vl}$
that depends on the gamma-ray line energy $l$. The material composition
and density conversions needed for the simulation are retrieved from
the CT scan using the conversion scheme of \cite{Schneider00}, whereas
the attenuation coefficients are extracted from the XCOM database
\parencite{nist_xcom}. These will be the input data for the next
simulation steps. Note that the model relies to some extent on prior
information of the initial CT scan, but can later be adjusted by leaving
the oxygen and carbon concentrations as free parameters $\rho^{vt}$.
The reason for focusing on those two elements is that they are the
most abundant nuclear targets for prompt gamma-ray emission in human
tissue \parencite{verburg_simu}. Other elements have a minor contribution
to the proton-induced signal, or none in the case of hydrogen.

\subsection{Proton energy spectrum\label{sec:ProtonEnergySpectrum}}

The stopping of the pencil-beams within the patient or phantom is
modeled using an extended version of gPMC \parencite{jia_gpmc,qin_gpumc},
which is a CUDA GPU-accelerated Monte Carlo code specifically designed
for proton therapy simulation. The simulations are performed on a
Tesla K40 GPU accelerator (NVIDIA Corporation, Santa Clara, CA). The
Monte Carlo beam model has been calibrated to match experimentally
measured Bragg peaks following the method of \textcite{verburg_mcauto}.
Through the use of the Monte Carlo method, geometries with complex
tissue inhomogeneities, range straggling and strong multiple Coulomb
scattering are modeled accurately.

To quantify the absolute proton range, we simulate several range error
scenarios $c$ in steps of 1 mm. The range of the protons in the GPU
Monte Carlo simulation is altered by slightly changing the incident
proton energy. We have automated the Monte Carlo simulation based
on the treatment plan, the converted densities $\hat{\rho}^{vt}$
and material compositions $m^{v}$. The output of the simulation is
the deposited dose and the proton energy spectrum, i.e., the number
of protons $N_{\text{p}}^{csve}$ with kinetic energy $e$ in each
FOV voxel $v$ for every pencil-beam spot $s$ and range error scenario
$c$.

\subsection{Prompt gamma-ray emission}

The next step involves the calculation of the number of prompt gamma-rays
with discrete energies $l$ emitted along the beam path. We do not
rely on a Monte Carlo simulation of the nuclear reactions and gamma-ray
emissions; only experimentally determined cross sections are used.
Besides the output data from the GPU Monte Carlo simulation $N_{\text{p}}^{csve}$
and the target nuclei concentrations $\hat{\rho}^{vt}$ from the CT
scan, the differential cross sections $\sigma^{elt}$ of every prompt
gamma-ray transition $l$ and target nucleus $t$ are needed, which
depend on the proton energy $e$. For the determination of the number
of prompt gamma-rays $N_{\gamma}^{csvlt}$ emitted along the beam
path for each pencil-beam $s$, range error scenario $c$, voxel $v$,
transition $l$ and target nucleus $t$, the voxel-dependent proton
energy distributions from the GPU Monte Carlo simulation are multiplied
with the energy-dependent cross sections and target concentrations:
\begin{equation}
N_{\gamma}^{csvlt}=\sum_{e}N_{\text{p}}^{csve}\cdot\frac{\hat{\rho}^{vt}N_{\mathrm{A}}}{A^{t}}\cdot4\pi\sigma^{elt},\label{eq:gammaemission}
\end{equation}
where $A^{t}$ is the molar atomic mass of target element $t$ and
$N_{\mathrm{A}}$ is the Avogadro constant. The lines included in
the model are described in \tabref{overviewIndices}, and correspond
to the nuclear reaction channels listed in \tabref{channels}.

\begin{table}
\caption{Nuclear reaction channels leading to abundant prompt gamma-ray emission
in human tissue \parencite{verburg_spec}, classified according to
the oxygen and carbon target nuclei. The index $l$ corresponds to
that of \tabref{overviewIndices} and the energies of the prompt gamma-rays
are given in MeV. Reactions which result in the emission of prompt
gamma-rays with similar energies are clustered at the same index $l$
as they cannot be resolved experimentally. \label{tab:channels} }

\centering{}{\small{}\small{}%
\begin{tabular}{llll}
\hline 
\textbf{Index $l$}  & \textbf{$\boldsymbol{\gamma}$ energies / MeV} & \textbf{Reactions with }\textsuperscript{\textbf{16}}\textbf{O} & \textbf{Reactions with }\textsuperscript{\textbf{12}}\textbf{C}\tabularnewline
\hline 
0  & 1.63  & $^{16}\text{O}\,(p,x\,\gamma)\,^{14}\text{N}$  & \tabularnewline
\hline 
1  & 2.00  & $^{16}\text{O}\,(p,x\,\gamma)\,^{11}\text{C}$  & $^{12}\text{C}\,(p,x\,\gamma)\,^{11}\text{C}$\tabularnewline
 & 2.04  & $^{16}\text{O}\,(p,x\,\gamma)\,^{15}\text{O}$  & \tabularnewline
\hline 
2  & 2.31  & $^{16}\text{O}\,(p,x\,\gamma)\,^{14}\text{N}$  & \tabularnewline
\hline 
3  & 2.74  & $^{16}\text{O}\,(p,p'\,\gamma)\,^{16}\text{O}$  & \tabularnewline
 & 2.79  & $^{16}\text{O}\,(p,x\,\gamma)\,^{14}\text{N}$  & \tabularnewline
 & 2.80  & $^{16}\text{O}\,(p,x\,\gamma)\,^{11}\text{C}$  & $^{12}\text{C}\,(p,x\,\gamma)\,^{11}\text{C}$\tabularnewline
 & 2.87  & $^{16}\text{O}\,(p,x\,\gamma)\,^{10}\text{B}$  & $^{12}\text{C}\,(p,x\,\gamma)\,^{10}\text{B}$ \tabularnewline
\hline 
4  & 4.44  & $^{16}\text{O}\,(p,x\,\gamma)\,^{12}\text{C}$  & $^{12}\text{C}\,(p,p'\,\gamma)\,^{12}\text{C}$\tabularnewline
 & 4.45 & $^{16}\text{O}\,(p,x\,\gamma)\,^{11}\text{B}$  & $^{12}\text{C}\,(p,x\,\gamma)\,^{11}\text{B}$ \tabularnewline
\hline 
5  & 5.24  & $^{16}\text{O}\,(p,x\,\gamma)\,^{15}\text{O}$  & \tabularnewline
 & 5.27  & $^{16}\text{O}\,(p,p\,p\,\gamma)\,^{15}\text{N}$  & \tabularnewline
 & 5.18  & $^{16}\text{O}\,(p,x\,\gamma)\,^{15}\text{O}$  & \tabularnewline
 & 5.30  & $^{16}\text{O}\,(p,p\,p\,\gamma)\,^{15}\text{N}$  & \tabularnewline
\hline 
6  & 6.13  & $^{16}\text{O}\,(p,p'\,\gamma)\,^{16}\text{O}$  & \tabularnewline
 & 6.18  & $^{16}\text{O}\,(p,x\,\gamma)\,^{15}\text{O}$  & \tabularnewline
\hline 
\end{tabular}{\small{}}}{\small\par}
\end{table}

In addition, the proton continuum spectrum is considered, as it contains
a significant fraction of the collected events. This continuum is
in part due to gamma-rays from the main discrete lines that do not
deposit their full energy in the detectors. Another part, which contributes
approximately to an equal extent, is the prompt gamma-ray quasi-continuum.
These are emissions of unresolved gamma-rays $\widetilde{N}_{\gamma}^{csvlta}$
from unknown transitions $l$ which are not resolvable in the measurement.
This part is approximated by a linear combination of known lines with
an effective cross section $\psi^{eta}$ depending on the proton energy
$e$, target $t$, and are associated with absorption processes $a=3,4$
($\psi^{eta}=0$ if $a<3$):

\begin{equation}
\widetilde{N}_{\gamma}^{csvlta}=\sum_{e}u^{l}N_{\text{p}}^{csve}\cdot\frac{\hat{\rho}^{vt}N_{\mathrm{A}}}{A^{t}}\cdot4\pi\psi^{eta},
\end{equation}
where the constants $u^{l}$ are normalized weights chosen such that
the continuum gamma-ray emission energy spectrum is approximately
constant up to 6 MeV.

The proton energy range that we model for the prompt gamma-ray emission
within the detector FOV ranges from 0 to 150 MeV. The incident energy
of the proton beam can be as high as the maximum energy of 230 MeV,
because the energy of the protons is reduced to well below 150 MeV
before they reach the FOV.

\subsection{Gamma-ray attenuation}

The calculation of the detection probability of the emitted gamma-rays
is separated into two parts: a measurement-specific calculation of
the gamma-ray attenuation based on the CT scan of the patient or phantom,
and a model of the collimator-detector system that is the same for
all measurements, which is discussed in the next section.

We define the transmission probability $\tau^{vl\theta\phi}$ for
a gamma-ray with energy $l$ traveling from FOV voxel $v$ to the
detector in the direction with angles $\left(\theta,\phi\right)$
that reaches the detector front plane without undergoing any interaction,
see \figref{fovBeamPerspective}. It is modeled using a ray tracing
algorithm \parencite{raytracing} with the energy-dependent attenuation
coefficients of the FOV voxels $\mu^{v'l}$ calculated previously.
The transmission probability is given by the sum along the path $\ell'$
that connects voxel $v$ with the crossing point on the front face
plane, that is defined by $\theta$ and $\phi$:
\begin{equation}
\tau^{vl\theta\phi}=\exp\left(-\sum_{v'}\mu^{v'l}\delta\ell'\right),
\end{equation}
where $\delta\ell'$ is the intersection length of the path $\ell'$
within the voxel $v'$.

\subsection{Collimator-detector system response}

The detection probability $\eta^{p\theta\phi lda}$ of the collimator-scintillator
system is determined using a Monte Carlo simulation of the electromagnetic
interactions using the TOPAS Monte Carlo code \parencite{TOPAS12}
with the Geant4 10.04.p01 toolkit \parencite{ALLISON2016186}. The
positions, dimensions and material compositions of all parts influencing
the detection probability, i.e., the collimator and scintillation
crystals, are included in the model. Photons with energies corresponding
to the prompt gamma-ray lines $l$ are simulated, originating from
the front plane of the detection system, which is divided into pixels
$p$ as shown in \figref{fovDetectorPerspective}. The simulated photons
are emitted isotropically from all front plane positions and directions
defined by angles $\theta$ and $\phi$. Five types of events corresponding
to different types of interactions $a$ in the detectors $d$ are
counted. Events in which the whole initial photon energy is deposited
in a single scintillation crystal ($a=0$) are counted as full energy
detection events, (1) single and (2) double escape lines are simulated
by counting the events where the deposited energy corresponds to the
initial photon energy reduced by 511\,keV, respectively 1.022\,MeV.
Additionally, any event with (3) a deposited energy between 1.4\,MeV
and 3.0\,MeV or (4) a deposited energy higher than 3.0\,MeV is recorded.

The total detection probability $\xi^{vlda}$ for a gamma-ray of transition
$l$ originating from voxel $v$, not interacting until reaching the
front plane and undergoing an absorption type $a$ in detector $d$
yields:
\begin{equation}
\xi^{vlda}=\sum_{\theta,\phi}\tau^{vl\theta\phi}\cdot\eta^{p\theta\phi lda},
\end{equation}
where $p$ is the pixel on the front plane that is intersected by
the path from voxel $v$ in the direction $\left(\theta,\phi\right)$.

\subsection{Absolute range verification\label{sec:AbsoluteRangeVerification}}

The expected counts $\zeta_{\gamma}^{csltda}$ according to the simulated
model are calculated for every range error scenario $c$, pencil-beam
spot $s$ of the treatment plan, target nucleus $t$, prompt gamma-ray
line $l$, interaction type $a$ and detector $d$:
\begin{equation}
\zeta_{\gamma}^{csltda}=\sum_{v}\left(N_{\gamma}^{csvlt}+\widetilde{N}_{\gamma}^{csvlta}\right)\cdot\xi^{vlda},\label{eq:modeled}
\end{equation}
and we sum over each detector row $r$:

\begin{equation}
\zeta_{\gamma}^{csltra}=\sum_{d=4r}^{4r+3}\zeta_{\gamma}^{csltda}.\label{eq:modeledrow}
\end{equation}

The results of the independent simulations with discrete range error
scenarios $c$ are interpolated to get a continuous model of the counts
($\zeta_{\gamma}^{csltra}\rightarrow\zeta_{\gamma}^{\epsilon sltra}$)
as a function of the absolute range error $\epsilon=(\check{c}-8)\times\left(1\,\text{{mm}}\right)$,
where $\check{c}$ is the interpolated counterpart of $c$.

The absolute proton range of a pencil-beam spot $s$ is determined
by minimizing the least square residuals between the experimentally
measured gamma-ray counts $C_{\gamma}^{slra}$ and the modeled ones
$\zeta_{\gamma}^{\epsilon sltra}$, leaving the range error scenario
$c$ and actual target concentrations $\rho^{vt}=k^{vt}\hat{\rho}^{vt}$
as free parameters:

\begin{equation}
\underset{\epsilon^{s},k^{s0},k^{s1},\chi^{s3}}{\mathrm{argmin}}\left(\sum_{l,r,a}w^{a}\left\Vert \chi^{sa}\sum_{t}k^{st}\zeta_{\gamma}^{\epsilon^{s}sltra}-C_{\gamma}^{slra}\right\Vert _{2}^{2}\right).\label{eq:RangeOptimization}
\end{equation}

In total, for every spot, four parameters are fitted: $\epsilon{}^{s},k^{s0},k^{s1},\chi^{s3}$.
The correction factors $k^{v0}$ and $k^{v1}$ of the target concentrations
($\rho^{vt}=k^{vt}\hat{\rho}^{vt}$) are assumed to be a constant
multiplier ($k^{vt}\rightarrow k^{st}$) within the region of interest
of the FOV of spot $s$. $\chi^{sa}$ is a multiplication factor that
is $1$ if $a<3$ and is left as free parameter for the continuum
absorption processes ($\chi^{s3}$=$\chi^{s4}$). This multiplier
allows to account for small concentrations of other nuclei that are
not modeled in the continuum emission, as well as small errors that
are introduced by the fact that the gamma-ray attenuation in the patient
cannot be exactly modeled for the continuum component. $w^{a}$ is
an empirical scaling constant which is 1 for $a<3$ and 0.1 otherwise,
to equalize the absolute contribution of each absorption process to
the least squares.

The measurement of the proton continuum $C_{\gamma}^{slr3}$ and $C_{\gamma}^{slr4}$
cannot be distinguished experimentally depending on $l$. The experimental
value corresponds to the sum of all lines, and is redistributed among
each line $l$ according to the weights $u^{l}$, and is compared
then to the values predicted by the simulation at absorption process
$a=3$ or $a=4$ and line $l$.

The Levenberg–Marquardt algorithm was used to solve the optimization
problem in \eqref{RangeOptimization}. For the present detector geometry,
the problem is convex and the optimization converges to the same global
optimum independently of the initial values of the parameters.

\subsection{Distal spot aggregation\label{sec:DistalSpotAggregation}}

Pencil-beam spots $\widetilde{s}$ that are delivered to the same
lateral position of spot $s$, but to an adjacent energy layer with
a proton range that differs by less than 10 mm, will be subject to
almost the same range error scenario $\epsilon^{s}\approx\epsilon^{\tilde{s}}$
and elemental concentration correction factors $k^{st}\approx k^{\tilde{s}t}$,
$\chi^{sa}\approx\chi^{\tilde{s}a}$. Instead of treating these spots
as being fully independent, it is useful to determine their range
error simultaneously to improve statistical precision. This is accomplished
by simultaneously fitting the parameters for these spots, rather than
by merging the histograms as done in the lateral direction (\secref{LateralMerging}).
Hence, for each spot $s$, we optimize:

\begin{equation}
\underset{\epsilon^{s},k^{s0},k^{s1},\chi^{s3}}{\mathrm{argmin}}\left(\sum_{s'}\sum_{l,r,a}w^{a}\left\Vert \chi^{sa}\sum_{t}k^{st}\zeta_{\gamma}^{\epsilon^{s}s'ltra}-C_{\gamma}^{s'lra}\right\Vert _{2}^{2}\right),\label{eq:RangeOptimizationMerged}
\end{equation}
where $s'$ is a set of spots including the pencil-beam $s$ and the
aggregated ones $\widetilde{s}$. When verifying the most distal energy
layer, this normally results in only one spot $\widetilde{s}$ from
the second most distal energy layer to be aggregated to the spot $s$
under consideration.

\subsection{Cross section optimization\label{sec:CrossSectionOptimization}}

The nuclear reaction cross sections $\sigma^{elt}$ for the prompt
gamma-ray production are the cornerstone of our model, which enable
the absolute range verification based on fundamental physical principles.
We previously published the cross sections that we use for \textsuperscript{16}O
and \textsuperscript{12}C nuclei for a 90 degree angle between the
incident beam and the detector \parencite{verburg_spec}. These are
based on a combination of our own measurements and previous literature
referenced therein. Unlike other measurements that are typically performed
at a few specific proton energies, we optimized the cross sections
for the complete proton energy range from 0 to 150 MeV, which is required
for range verification of beams stopping in matter. The range verification
is not sensitive to small resonances in the cross sections because
of the energy spread of the proton beam close to the end-of-range.

As is typical for cross section measurements, these are subject to
some systematic errors due to issues such as the bias in the separation
of the gamma-ray line from the background. Because our present system
features new data acquisition algorithms and improved methods to separate
the gamma-ray lines from the background, we have re-optimized our
previous cross sections $\widehat{\sigma}^{elt}$ by applying a small
correction based on a reference measurement in water and in high-density
polyethylene. Because of the improvements, we expect the re-optimized
cross sections to have a smaller systematic error as compared to our
previous optimization. By using the same system for the cross section
optimization and the range verification, this re-optimization is also
effectively a calibration that removes the effect of the system-specific
systematic bias.

To obtain the optimized cross sections $\sigma^{elt}$, we minimize
the discrepancy between the measured gamma-ray counts $C_{\gamma}^{slra}$
and the values predicted by the model $\zeta_{\gamma}^{csltra}$ for
the nominal case $c=8$ (0 mm range error) and each target $t=0,1$,
line $l$ and proton energy $e$:

\begin{equation}
\underset{\sigma^{elt}}{\mathrm{argmin}}\left(\sum_{a=0}^{2}\sum_{s,r}\left\Vert k^{st}\zeta_{\gamma}^{8sltra}-C_{\gamma}^{slra}\right\Vert _{2}^{2}+\lambda\left\Vert \sigma^{elt}-\hat{\sigma}^{elt}\right\Vert _{2}^{2}\right),\label{eq:csopt}
\end{equation}
where
\begin{equation}
\zeta_{\gamma}^{8sltra}=\sum_{d=4r}^{4r+3}\sum_{e}N_{\text{p}}^{8sve}\cdot\frac{\rho^{vt}N_{\mathrm{A}}}{A_{t}}\cdot4\pi\left(\sigma^{elt}+u^{l}\cdot\psi^{eta}\right)\cdot\xi^{vlda}
\end{equation}
 and $\lambda$ is an empirical damping factor. The first term is
the least squares optimization between the model and the measurement,
while the second one is a Tikhonov regularization term to penalize
large differences between the reference cross sections $\widehat{\sigma}^{elt}$
and the optimized ones $\sigma^{elt}$. Note that, in this calibration
experiment, $k^{st}$ is fixed as the target composition is well known.

Likewise, for the effective cross sections $\psi^{eta}$ that describe
the unresolved continuum, we optimize for $a=3$ and $a=4$:

\begin{equation}
\underset{\psi^{eta}}{\mathrm{argmin}}\left(\sum_{s,l,r}\left\Vert k^{st}\zeta_{\gamma}^{8sltra}-C_{\gamma}^{slra}\right\Vert _{2}^{2}\right).\label{eq:csoptTotal}
\end{equation}

\section{Experimental setups\label{sec:ExperimentalSetups}}

\subsection{Cross section optimization\label{sec:ExpSetupCrossSection}}

The measurements for the cross section optimization were performed
during the delivery of a single high-dose spot ($3\times10^{10}$
protons) along the central beam axis consecutively to 19 energy layers
to either a water phantom or to a solid block of polyethylene. The
phantom was mounted on a linear stage on breadboard for accurate positioning.
We estimate the positioning uncertainty in depth to be $\pm0.5\,\mathrm{mm}$.
The proton ranges were between 10 cm and 15 cm water equivalent, corresponding
to incident energies of 116 MeV to 145 MeV. In this setup, the detector
FOV (see \figref{fov}) was centered at a depth of $-x_{0}=124\,\text{{mm}}$
in the phantom and the distance between the isocenter and the front
plane of the collimator was 150 mm ($z_{\text{{iso}}}=10\,\text{{mm}}$),
including an air gap of 100 mm.

\subsection{Absolute range verification\label{sec:ExpSetupRangeVerification}}

\begin{table}
\caption{Characteristics of the treatment plan designed for delivering 0.9
Gy dose uniformly to a $5.3\times10\times10\,\text{cm}^{3}$ target
volume in water with a beam current of 2 nA, see \figref[s]{ExperimentalSetups}
and \ref{fig:ModelVisual}, divided in eight iso-energy layers. The
reported proton range\textbf{\small{} $R_{80}$} is the physical distance
from the phantom front face. \label{tab:tplan}}

\centering{}{\small{}\small{}%
\begin{tabular}{lllllll}
\textbf{\small{}Layer } & \textbf{\small{}Proton energy } & \textbf{\small{}Range $R_{80}$ } & \textbf{\small{}\# spots } & \textbf{\small{}\# protons} & \textbf{\small{}Protons / spot} & \textbf{\small{}Dose to target}\tabularnewline
 & MeV & mm &  & $\times10^{9}$ & $\times10^{9}$ (mean) & Gy (mean)\tabularnewline
\hline 
{\small{}0 } & {\small{}176 } & {\small{}210} & {\small{}\hphantom{0}203 } & {\small{}24.1 } & {\small{}0.12 } & {\small{}0.25 }\tabularnewline
{\small{}1 } & {\small{}173 } & {\small{}203 } & {\small{}\hphantom{0}217 } & {\small{}29.1 } & {\small{}0.13 } & {\small{}0.32 }\tabularnewline
{\small{}2 } & {\small{}168 } & {\small{}194 } & {\small{}\hphantom{0}199 } & {\small{}12.8 } & {\small{}0.06 } & {\small{}0.13 }\tabularnewline
{\small{}3 } & {\small{}164 } & {\small{}186 } & {\small{}\hphantom{0}180 } & {\small{}\hphantom{0}8.2 } & {\small{}0.05 } & {\small{}0.08 }\tabularnewline
{\small{}4 } & {\small{}160 } & {\small{}178 } & {\small{}\hphantom{0}180 } & {\small{}\hphantom{0}7.7 } & {\small{}0.04 } & {\small{}0.06 }\tabularnewline
{\small{}5 } & {\small{}156 } & {\small{}171 } & {\small{}\hphantom{0}167 } & {\small{}\hphantom{0}6.3 } & {\small{}0.04 } & {\small{}0.04 }\tabularnewline
{\small{}6 } & {\small{}152 } & {\small{}163 } & {\small{}\hphantom{0}127 } & {\small{}\hphantom{0}4.0 } & {\small{}0.03 } & {\small{}0.02 }\tabularnewline
{\small{}7 } & {\small{}148 } & {\small{}155 } & {\small{}\hphantom{0}137 } & {\small{}\hphantom{0}5.7 } & {\small{}0.04 } & {\small{}0.02 }\tabularnewline
\hline 
{\small{}Total} &  &  & {\small{}1410} & {\small{}97.9 } & {\small{}0.07 } & {\small{}0.91 }\tabularnewline
\end{tabular}{\small{}}}{\small\par}
\end{table}

\begin{figure}
\subfloat[\label{fig:setupnsh}Nominal]{\includegraphics[scale=0.65]{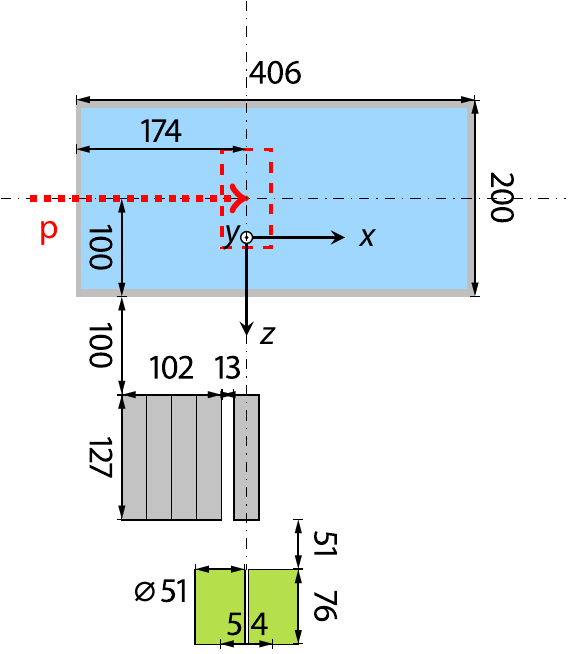}

}\hfill{}\subfloat[\label{fig:setuphsw}Solid water insert]{\includegraphics[scale=0.65]{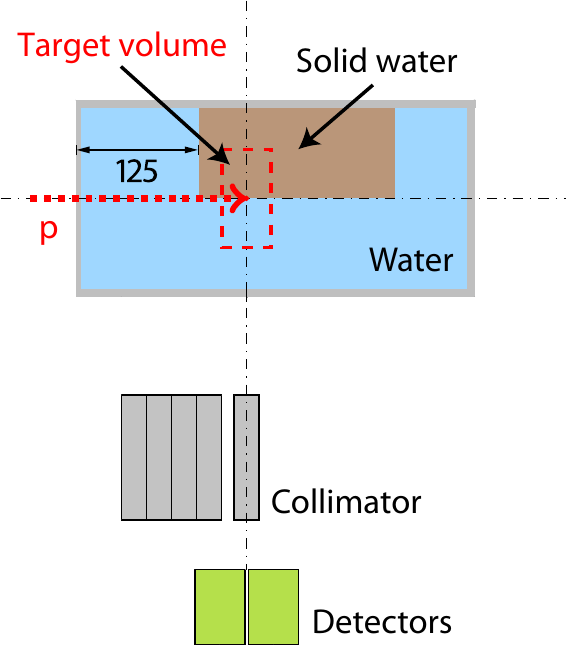}

}\hfill{}\subfloat[\label{fig:setupwsh}2.2; 5.2 mm shifter]{\includegraphics[scale=0.65]{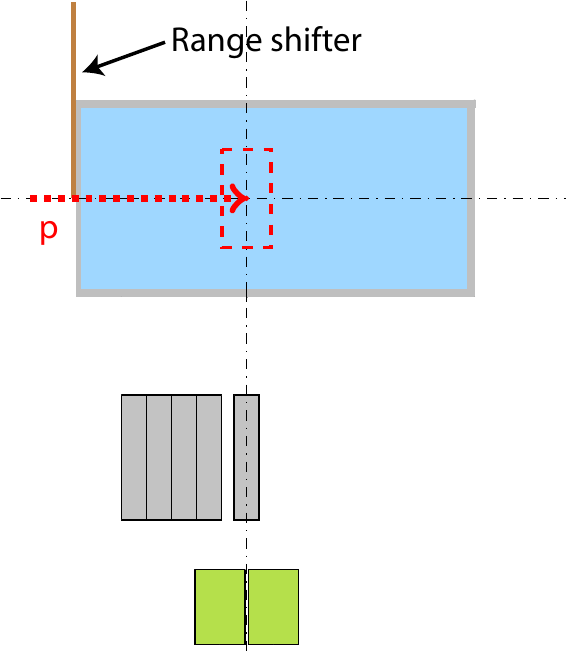}

}\hfill{}\subfloat[\label{fig:setupwbi}Inner; Cortical bone]{\includegraphics[scale=0.65]{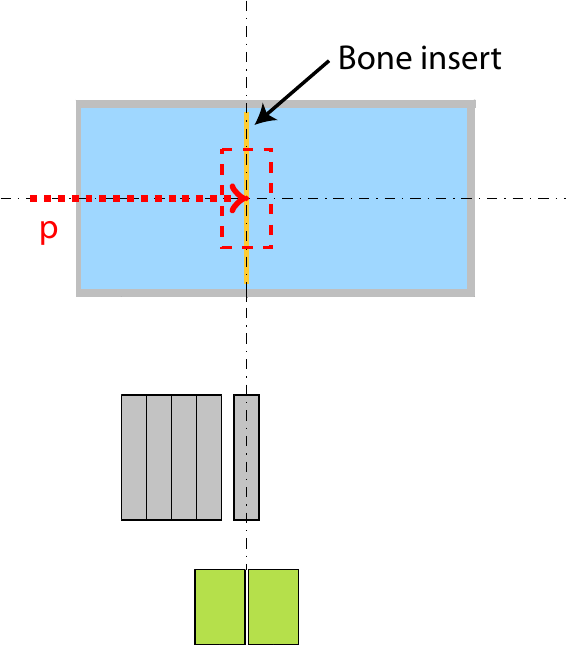}

}

\caption{Schematic of the experimental setups to assess the range verification
performance (top view, dimensions are in mm). Protons (red arrow)
irradiate the water phantom (blue: water, light gray: walls), where
$x_{0}=-174\,\text{{mm}}$ and $z_{\text{{iso}}}=-40$ mm. The FOV
coordinate system is defined in \figref{fov}. The black dash-dotted
lines intersect at isocenter. The dashed red rectangle is the target
volume ($5.3\times10\times10$ cm\protect\textsuperscript{3}). The
prompt gamma-rays are collimated (gray) and measured with scintillation
detectors (green). (a) Reference case in water with no range error.
(b) Solid water block (brown) covering half of the field. (c) 2.2
mm or 5.2 mm water equivalent range shifter (brown) covering half
of the field. (d) 5 mm thick slab (orange) of inner or SB3 cortical
bone inserted in the middle of the field. \label{fig:ExperimentalSetups}}
\end{figure}

\begin{figure}
\begin{centering}
\includegraphics[width=0.75\columnwidth]{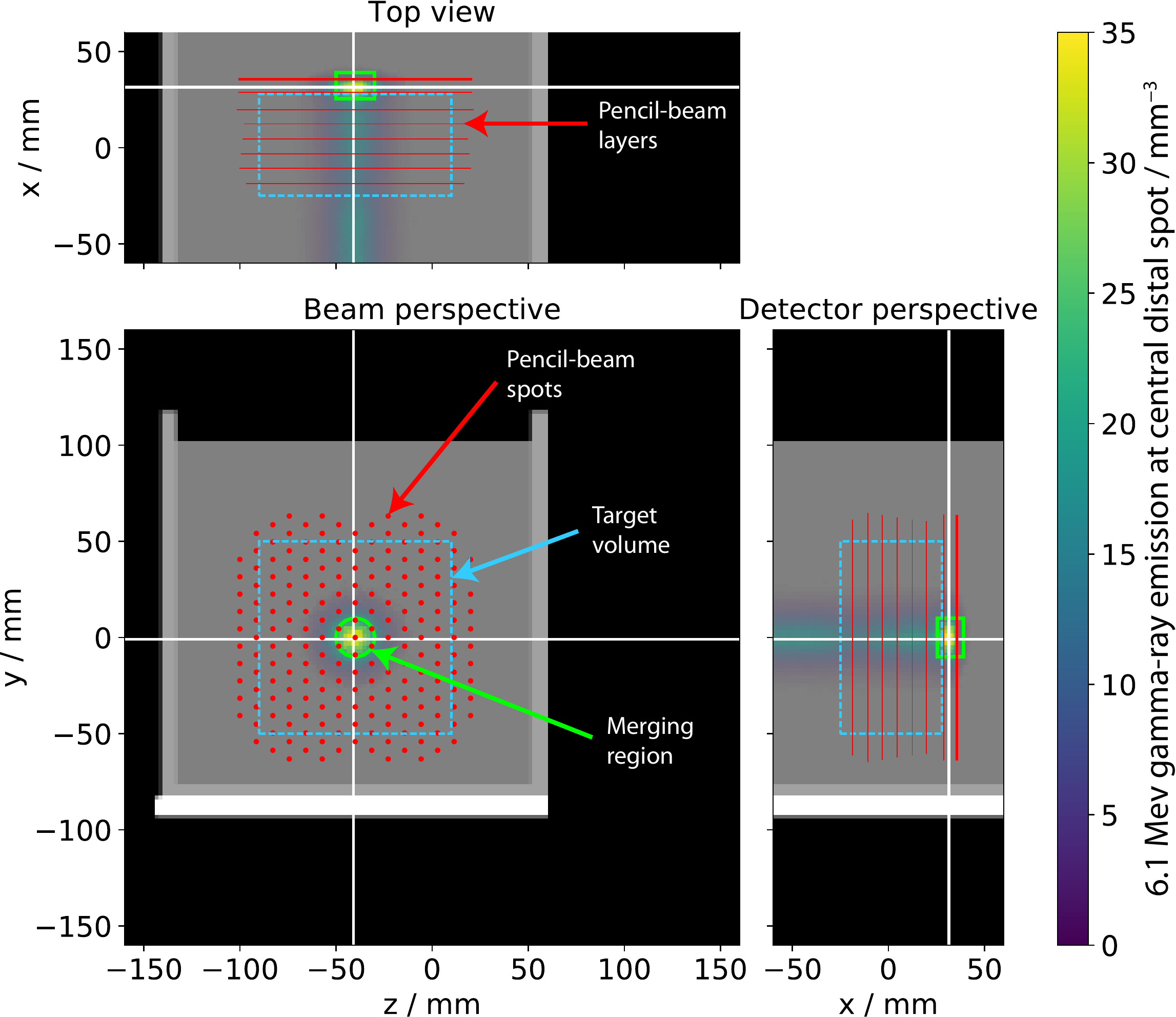}
\par\end{centering}
\textsf{\caption{Visualization of the planning images and pencil-beam positions in
the treatment plan, superimposed with the modeled 6.1 MeV prompt gamma-ray
emission density $N_{\gamma}^{c=8,s=101,v,l=6,t=0}/(1\times2\times2\,\text{mm}^{3})$
for the central spot of the distal layer, see \eqref{gammaemission}.
Relative to the field of view (\figref{fov}), the top view is at
slice $y=0$ mm, the beam perspective is at slice $x=31$ mm, and
the detector perspective is at slice $z=-40$ mm (white cross-hair
lines). The phantom setup corresponds to \figref{setupnsh}, placed
at $x_{0}=-174\,\text{{mm}}$ and $z_{\text{{iso}}}=-40$ mm. The
pencil-beam spot scanning pattern is overlaid in the beam's eye view,
as red spots. The green cylinder (circle or rectangle in the projections)
indicates a 10 mm merging radius and a 10 mm depth, enclosing spots
from the two distal energy layers, as explained in \secref[s]{LateralMerging}
and \ref{sec:DistalSpotAggregation}. The dashed blue rectangles represent
the target region of the treatment plan. The red lines mark the nominal
ranges $R_{80}$ of the pencil-beam energy layers (\tabref{tplan}).\textsf{\label{fig:ModelVisual} }}
}
\end{figure}

To test the range verification with a clinically realistic pencil-beam
field, we designed a treatment plan to deliver a uniform dose of 0.9
Gy to a $5.3\times10\times10\,\text{cm}^{3}$ target in water, as
shown in \figref{setupnsh}. The isocenter was located centrally in
the target and the distance between isocenter and the collimator front
face was 200 mm ($z_{\text{{iso}}}=-40$ mm). There was an air gap
of 100 mm between the collimator and the phantom, which would also
be realistic when the system is used clinically. The FOV center was
at a depth of $-x_{0}=174\,\text{{mm}}$ mm in the phantom, see \figref{fov}.

The treatment plan was created using our in-house Astroid treatment
planning system \parencite{kooy_astroid}. The dose was delivered
to the target volume with a total of 1410 pencil-beams in eight energy
layers, as described in \tabref{tplan}. The uniform dose region of
the proton pencil-beams extends between 15 cm and 20.3\,cm depth.
The lateral distance between two neighboring pencil-beam spots in
the distal energy layer is approximately 9\,mm and the separation
between consecutive energy layers is about 8 mm. We used the standard
clinical CT conversion, assuming the phantom to consist of soft tissue.
This means that the model has no prior knowledge of the actual elemental
composition, which is left as free parameter $k^{st}$ to be fitted
simultaneously to the absolute range error $\epsilon^{s}$, see \eqref{RangeOptimizationMerged}.

Figure \ref{fig:ModelVisual} shows the pencil-beam scanning treatment
plan and the prompt gamma-ray emission for one of the pencil-beam
spots. We also show the merging region that defines the area in which
we aggregate the measurement data for range verification, as discussed
in \secref[s]{LateralMerging} and \ref{sec:DistalSpotAggregation}.
Altogether, including the lateral merging, the merging region corresponding
to spot $s$ can be visualized as a cylinder of 10 mm radius and 10
mm depth, enclosing typically a total of 14 pencil-beams, i.e. 7 from
each layer. 

The range verification performance was further assessed by introducing
different inhomogeneities in this setup, as shown in \figref{ExperimentalSetups}:
\begin{itemize}
\item Setup \ref{fig:setupnsh} is the reference case where no range error
is expected.
\item In setup \ref{fig:setuphsw}, a solid water block (Gammex RMI, Middleton
WI) is inserted in half of the water phantom at a depth of 125 mm.
This setup is devoted to test the capability of the prototype to accurately
detect the absolute range independently of the elemental composition
of the irradiated tissue. Solid water has a stopping power close to
that of water (1.02 ratio), but a by-mass elemental composition of
67\% C, 20\% O and 8\% H \parencite{hunemohr_gammex}, very different
from that of water (89\% O, 11\% H). A shift of 1.7 mm of the end-of-range
is expected for the pencil-beams of the distal energy layer going
through the solid water.
\item Setup \ref{fig:setupwsh} includes a range shifter placed in front
of the water phantom and covering half of the field. We used range
shifters with a water equivalent thickness of 2.2 and 5.2 mm. 
\item In setup \ref{fig:setupwbi}, a bone equivalent slab with a thickness
of 5 mm of inner or SB3 cortical bone (Gammex) is inserted into the
water phantom, centered at a depth of 174 mm. The expected shifts
of the end-of-range are 0.5 mm for inner bone and 3.0 mm for cortical
bone.
\end{itemize}
The expected range shifts created by the Gammex materials were determined
based on proton stopping power measurements that we performed by measuring
the shift in the Bragg peak position in water when the materials are
placed in the beam path. The stopping powers that we measured are
consistent with the measurements by \textcite{saito_gammex,hunemohr_gammex}.

\section{Results\label{sec:Results}}

\begin{figure}
\begin{centering}
\includegraphics[width=1\textwidth]{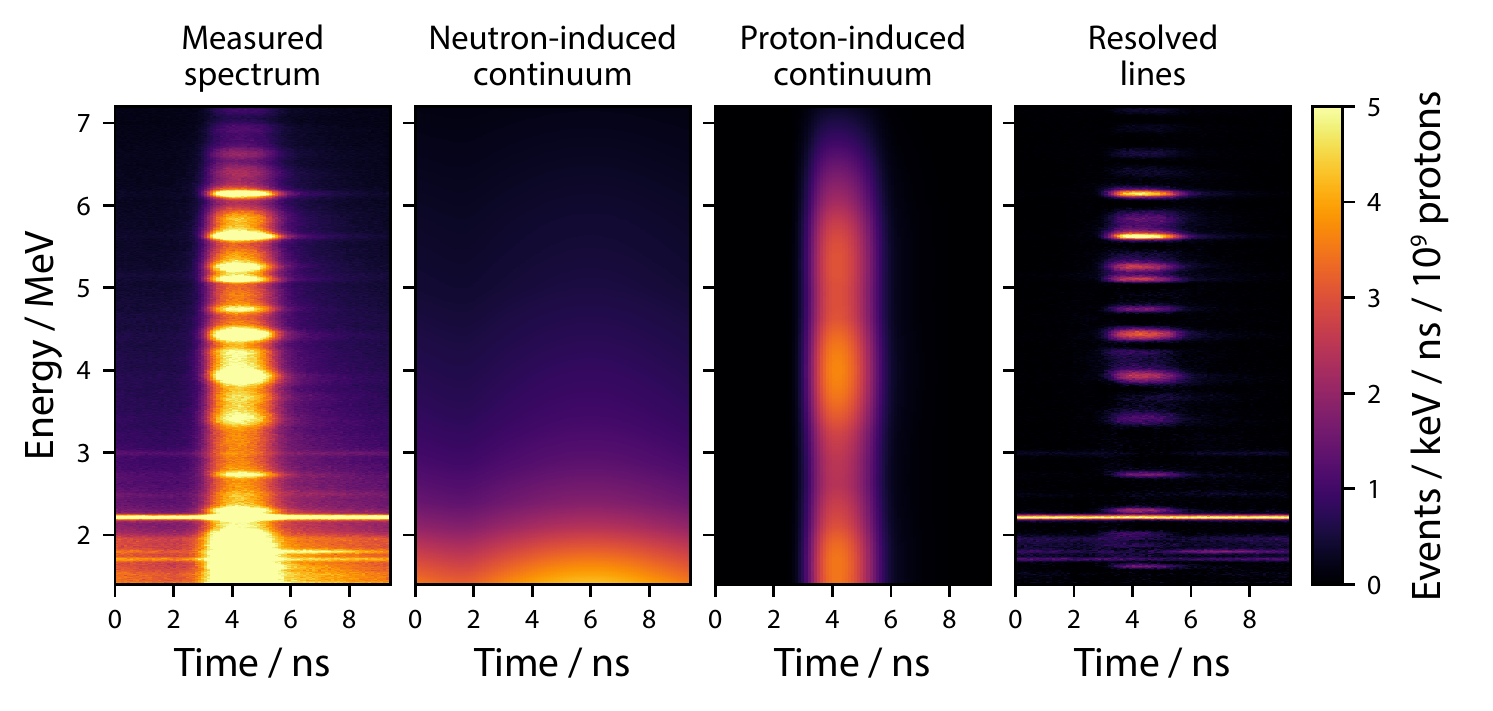}
\par\end{centering}
\caption{From left to right: measured 2D spectrum of the energy deposit over
trigger time with respect to the cyclotron radiofrequency; the neutron-
and proton-continuum background estimated by our algorithm; resolved
neutron-induced and proton-induced lines. This measurement was performed
during the irradiation of a water phantom with 2 nA beam current.
These spectra were measured by detector row $r=1$ during the cross
section optimization experiment (\secref{CrossSectionOptimization}),
where a high dose was delivered. Piled-up or coincident events have
been excluded.}
\label{fig:spectra}
\end{figure}

\subsection{Energy- and time-resolved histograms}

An example of the acquired energy- and time-resolved prompt gamma-ray
histogram and the separation into the different components (\ref{sec:HistogramAnalysis})
is presented in \figref{spectra}. This histogram was acquired during
the cross section calibration experiment (\secref{ExpSetupCrossSection}),
in which a high dose was delivered. It shows the excellent performance
of the detectors and the data acquisition system at a beam current
of 2 nA. During the delivery of the distal layer, the count rate in
each detector was on the order of $10^{6}$ events per second. Even
under these conditions, an energy resolution of 1.3\% full width at
half maximum was obtained at 6.1 MeV, allowing the gamma-ray lines
to be clearly resolved. A robust separation of the continuum and resolved
components can be observed. The background constitutes approximately
half of the detected gamma-rays, including sources such as the proton
treatment head, phantom, couch, walls and floor of the treatment room.
There is also some production of neutron-induced gamma-rays in the
collimator, but most of these are re-absorbed internally.

\subsection{Cross section optimization}

The consistency of the re-optimized cross sections was verified by
calculating the relative deviation $\Delta^{lt}=\langle1-\sigma^{elt}/\hat{\sigma}^{elt}\rangle$
between the optimized and our previous cross sections \parencite{verburg_spec},
averaged over the proton energy range from 0 MeV to 150 MeV. For the
more prominent prompt gamma-ray lines, the differences are $\Delta^{40}=17\%$,
$\Delta^{41}=5\%$, and $\Delta^{61}=12\%$.

\subsection{Absolute range verification}

\begin{figure}
\begin{tabular}{p{5.3cm}>{\raggedright}p{4.6cm}>{\raggedright}p{4.6cm}}
\subfloat[Nominal case, \figref{setupnsh}.]{\includegraphics[scale=0.9]{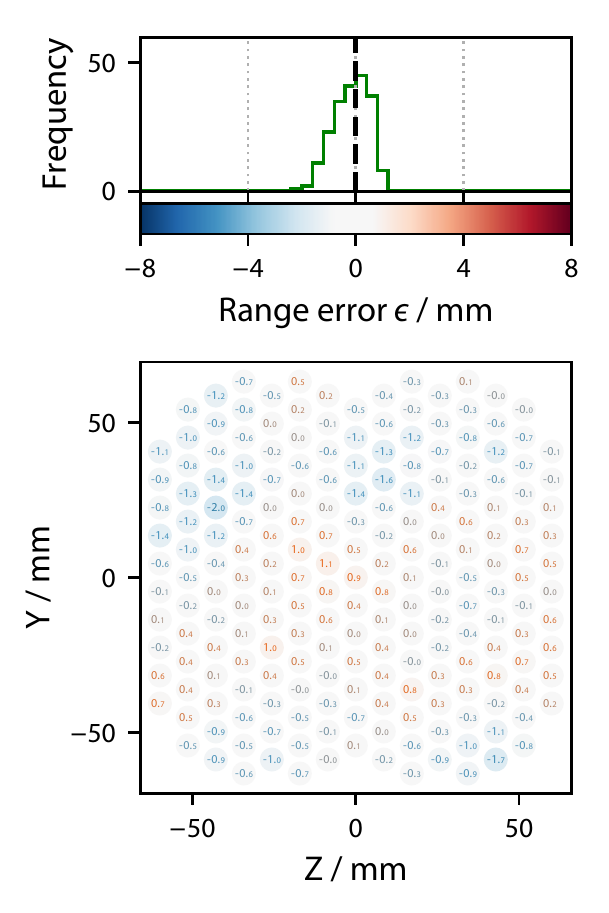}\label{fig:nsh}

}

\subfloat[Solid water insert on left, \figref{setuphsw}.]{\includegraphics[viewport=0bp 0bp 175bp 263bp,scale=0.9]{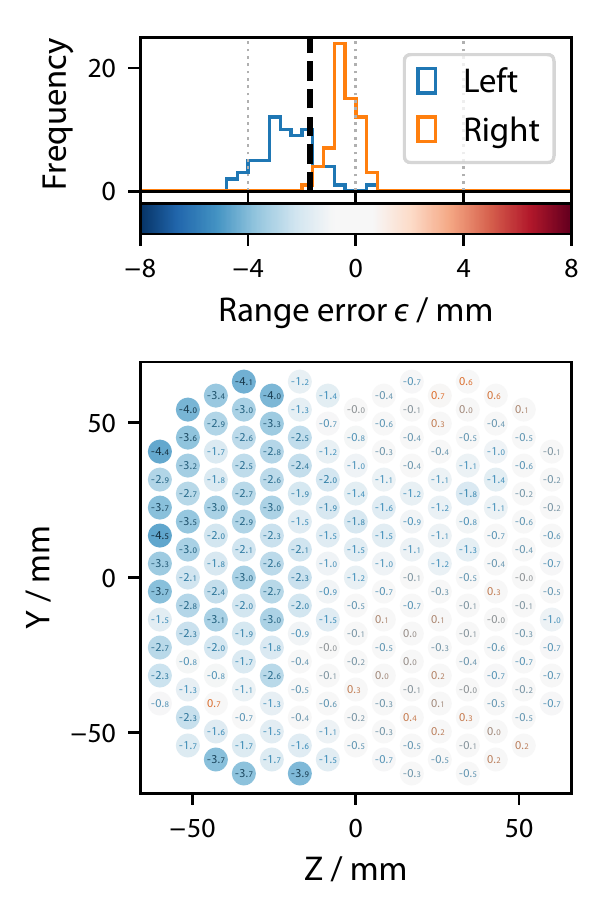}\label{fig:hsw}

} & \subfloat[2.2 mm shifter on left, \figref{setupwsh}.]{\includegraphics[viewport=0bp 0bp 157bp 263bp,scale=0.9]{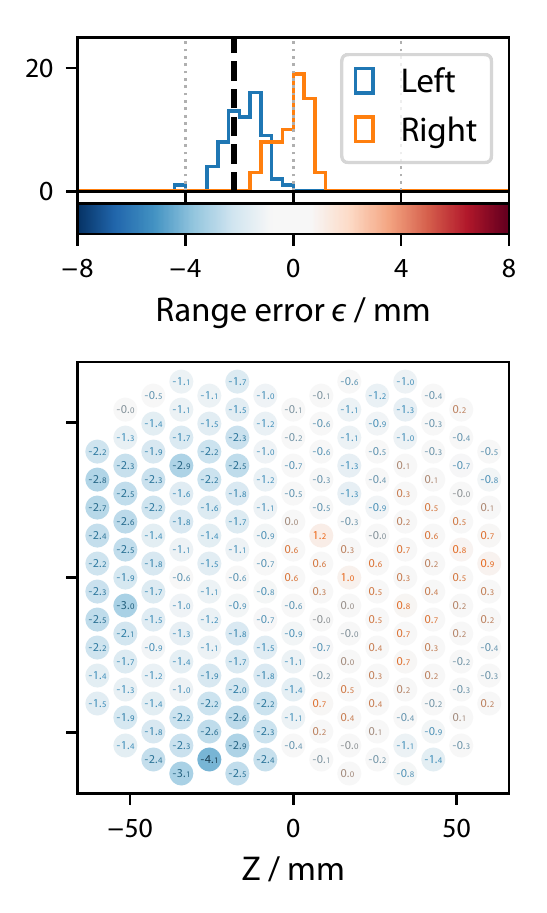}\label{fig:wsh22}

}

\subfloat[5.2 mm shifter on left, \figref{setupwsh}.]{\includegraphics[scale=0.9]{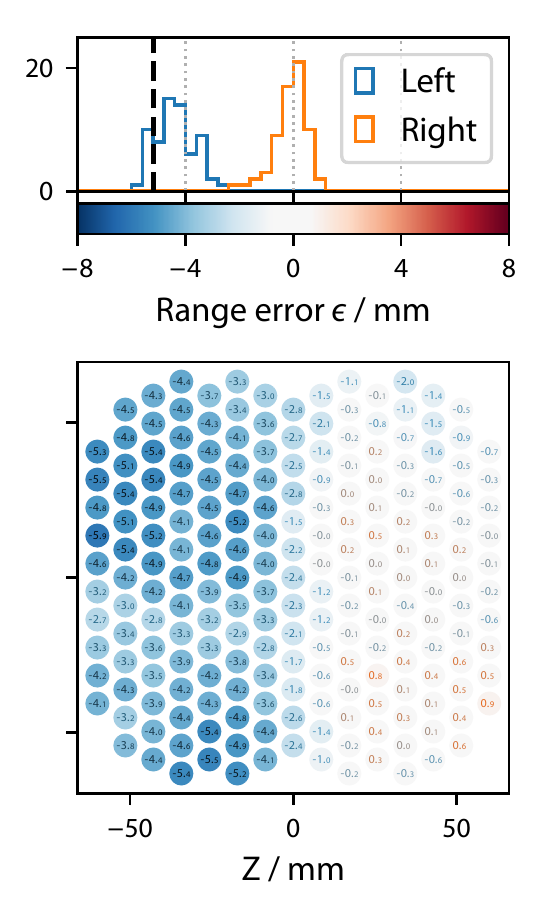}\label{fig:wsh52}

} & \subfloat[Inner bone, \figref{setupwbi}.]{\includegraphics[viewport=0bp 0bp 157bp 263bp,scale=0.9]{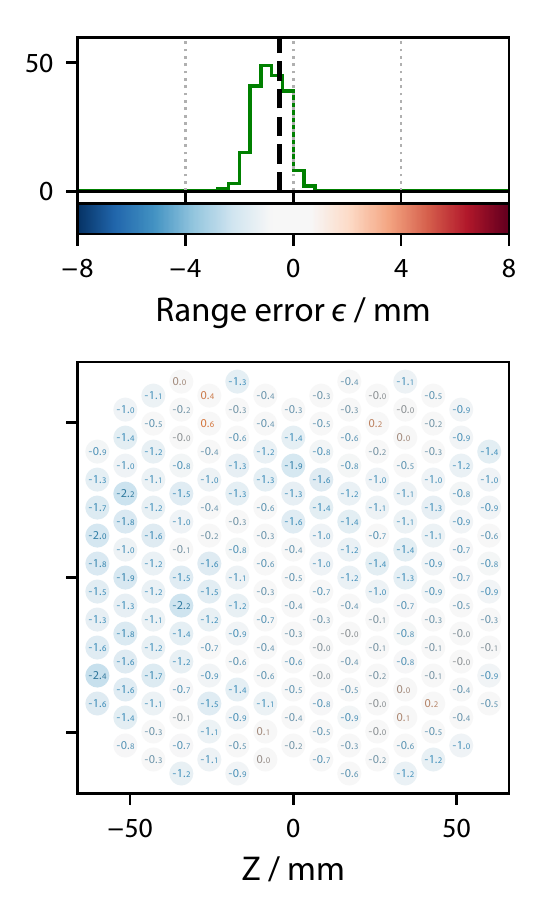}\label{fig:wibi}

}

\subfloat[Cortical bone, \figref{setupwbi}.]{\includegraphics[viewport=0bp 0bp 157bp 263bp,scale=0.9]{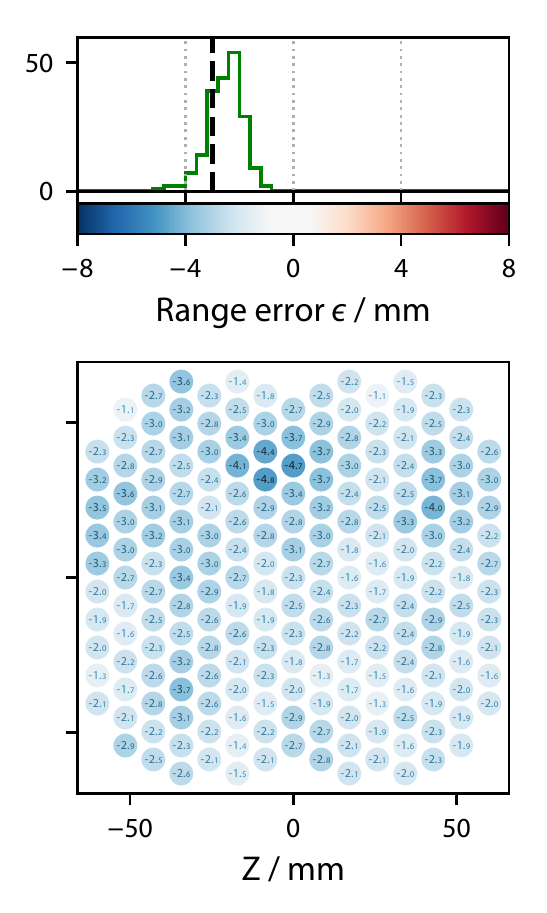}\label{fig:wcbi}

}\tabularnewline
\end{tabular}\caption{Range verification of the distal layer of a proton pencil-beam field
delivering 0.9 Gy to a $5.3\times10\times10\,\text{cm}^{3}$ region.
The $Y$ and $Z$ axis correspond to the $y$ and $z-z_{\mathrm{iso}}$
axes of \figref{fov}. The proton range error $\epsilon^{s}$ with
respect to the model prediction is annotated within the colored circles.
Shown are the experiments without range shifter (a), with a solid
water insert in the left half of the field (b), with a 2.2 mm (c)
and 5.2 mm (d) range shifter covering the left half of the field,
and with the inner (e) and SB3 cortical (f) bone inserts. Each spot
contains information from a cylindrical merging region of 10 mm depth
and 10 mm radius (\figref{ModelVisual}). The histograms show the
range errors with a bin width of 0.4 mm, in which a dashed black vertical
line marks the theoretically introduced range shift. Where relevant,
the histograms are shown separately for the left ($Z<-20$ mm) and
right ($Z>20$ mm) parts of the field.}
\label{fig:nsh-hsw-wsh-wbi} 
\end{figure}

\begin{figure}
\begin{centering}
\subfloat[Nominal case with liquid water, see \figref{setupnsh}.]{\centering{}\includegraphics[clip,scale=0.95]{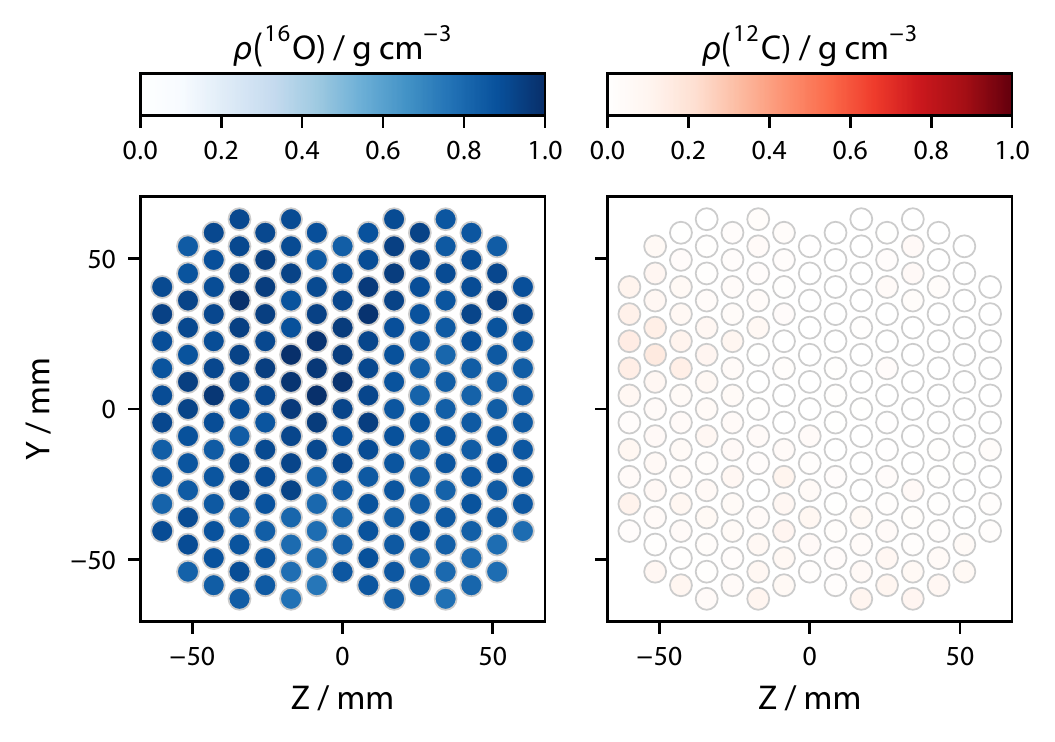}\emph{}\label{fig:nshC}}
\par\end{centering}
\begin{centering}
\subfloat[Solid water insert on the left half of the field, see \figref{setuphsw}.
\label{fig:hswC}]{\noindent \centering{}\includegraphics[clip,scale=0.95]{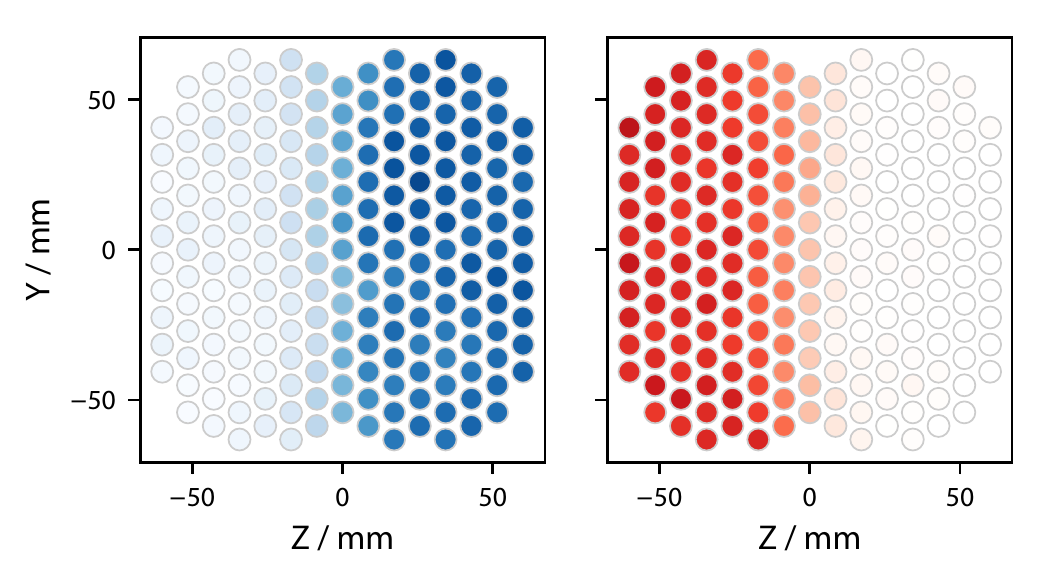}}
\par\end{centering}
\caption{Elemental composition determination for the same pencil-beam field
as in \figref{nsh-hsw-wsh-wbi}. The oxygen and carbon concentrations
by mass $\rho^{st}$ are shown in the left and right columns. The
reference case (a) is compared with the solid water insert in the
left half of the field (b). Each spot contains information from a
cylindrical merging region of 10 mm depth and 10 mm radius.\label{fig:nsh-hsw} }
\end{figure}

\begin{table}
\caption{Determined range error and elemental concentrations for the pencil-beams
irradiating the different setups, averaged over the distal energy
layer. The mean absolute range error and the mean statistical precision
($\pm2\sigma$) are listed. The range error $\epsilon$ refers to
the difference between the measured range of the pencil-beams and
the model prediction based on the treatment plan. Where relevant,
the results are given separately for the left part of the field ($Z<-20$
mm) and the right part ($Z>20$ mm). The statistics were calculated
based on 12 repeated measurements. \label{tab:RangeResults}}

\centering{}{\small{}\footnotesize{}%
\begin{tabular}{l|>{\raggedright}p{0.1025\columnwidth}>{\raggedright}p{0.1025\columnwidth}|>{\raggedright}p{0.1075\columnwidth}>{\raggedright}p{0.1075\columnwidth}|>{\raggedright}p{0.1075\columnwidth}>{\raggedright}p{0.1075\columnwidth}}
\hline 
\textbf{Setup } & \multicolumn{2}{l|}{\textbf{Range error $\mathbf{\epsilon}$ / mm}} & \multicolumn{2}{l|}{\textbf{$\mathbf{\rho(^{16}\mathrm{O})}$ }/\textbf{ g cm$^{-3}$}} & \multicolumn{2}{l}{$\mathbf{\rho(^{12}\mathrm{C})}$ /\textbf{ g cm$^{\boldsymbol{-3}}$}}\tabularnewline
 & \emph{Left}  & \emph{Right}  & \emph{Left}  & \emph{Right} & \emph{Left}  & \emph{Right} \tabularnewline
\hline 
\ref{fig:setupnsh} - Nominal & \multicolumn{2}{l|}{\hspace{2em}$-0.2\pm1.0$} & \multicolumn{2}{l|}{\hspace{2em}$0.88\pm0.09$} & \multicolumn{2}{l}{\hspace{2em}$0.05\pm0.06$}\tabularnewline
Expected & \multicolumn{2}{l|}{\hspace{2em}$\hphantom{{-}}0.0$} & \multicolumn{2}{l|}{\hspace{2em}$0.89$} & \multicolumn{2}{l}{\hspace{2em}$0.00$}\tabularnewline
\hline 
\ref{fig:setuphsw} - Solid water & $-2.4\pm1.9$ & $-0.5\pm1.0$  & $0.17\pm0.05$  & $0.88\pm0.10$ & $0.76\pm0.08$ & $0.03\pm0.05$\tabularnewline
Expected & $-1.7$  & $\hphantom{{-}}0.0$  & $0.20$ & $0.89$ & $0.67$ & $0.00$\tabularnewline
\hline 
\ref{fig:setupwsh} - 2.2 mm shifter & $-1.9\pm1.1$  & $-0.4\pm0.9$  & $0.90\pm0.10$ & $0.88\pm0.07$ & $0.07\pm0.08$ & $0.03\pm0.05$\tabularnewline
Expected & $-2.2$  & $\hphantom{{-}}0.0$  & $0.89$ & $0.89$ & $0.00$ & $0.00$\tabularnewline
\hline 
\ref{fig:setupwsh} - 5.2 mm shifter & $-4.4\pm1.1$  & $-0.4\pm0.9$  & $0.88\pm0.09$ & $0.88\pm0.07$ & $0.07\pm0.08$ & $0.02\pm0.05$\tabularnewline
Expected & $-5.2$  & $\hphantom{{-}}0.0$  & $0.89$ & $0.89$ & $0.00$ & $0.00$\tabularnewline
\hline 
\ref{fig:setupwbi} - Inner bone & \multicolumn{2}{l|}{\hspace{2em}$-0.8\pm1.0$} & \multicolumn{2}{l|}{\hspace{2em}$0.87\pm0.10$} & \multicolumn{2}{l}{\hspace{2em}$0.09\pm0.08$}\tabularnewline
Expected & \multicolumn{2}{l|}{\hspace{2em}$-0.5$} & \multicolumn{2}{l|}{\hspace{2em}$0.89$} & \multicolumn{2}{l}{\hspace{2em}$0.00$}\tabularnewline
\hline 
\ref{fig:setupwbi} - Cortical bone  & \multicolumn{2}{l|}{\hspace{2em}$-2.3\pm1.0$} & \multicolumn{2}{l|}{\hspace{2em}$0.85\pm0.09$} & \multicolumn{2}{l}{\hspace{2em}$0.10\pm0.08$}\tabularnewline
Expected & \multicolumn{2}{l|}{\hspace{2em}$-3.0$} & \multicolumn{2}{l|}{\hspace{2em}$0.89$} & \multicolumn{2}{l}{\hspace{2em}$0.00$}\tabularnewline
\hline 
\end{tabular}{\small{}}}{\small\par}
\end{table}

In \figref{nsh-hsw-wsh-wbi}, the absolute range verification is shown
for the experiments in which a dose of 0.9 Gy was delivered to a target
in different phantom setups. These results were obtained for each
spot of the most distal energy layer, using a cylindrical merging
region of 10 mm radius and 10 mm depth, as shown in \figref{ModelVisual}.
Note that the range errors are not calculated with respect to a previous
measurement, but with respect to the value predicted by the simulated
model with only the planning CT as prior information.

The experimental setups correspond to \figref{ExperimentalSetups}:
featuring no range shifter (\figref{setupnsh}), a solid water insert
in half of the field (\figref{setuphsw}), a range shifter covering
half of the field upstream of the water phantom (\figref{setupwsh})
and a bone insert slab within the phantom covering the full field
(\figref{setupwbi}). 

In \figref{nsh-hsw}, the determined oxygen and carbon concentrations
are shown for two of the experiments: the nominal case without range
shifter (\figref{setupnsh}) and the setup with a solid water insert
in half of the field (\figref{setuphsw}). The differences in elemental
composition in the different parts of the field can be clearly identified.

To determine the mean accuracy and the mean statistical precision
of the range verification, we repeated all of the above experiments
12 times. The results are listed in \tabref{RangeResults}. Averaging
all experiments, the mean systematic deviation in the determined range
was 0.5 mm. The mean statistical precision, determined by calculating
the standard deviation of the repeat measurements, was 1.1 mm at a
95\% confidence level ($2\sigma$) for the chosen merging radius of
10 mm and 10 mm depth. For the case of the solid water insert (\figref{hsw}),
we obtain a larger statistical uncertainty (1.9 mm) in the left part
of the field. This is in accordance with the lower prompt gamma-ray
yield of carbon with respect to oxygen targets \parencites{verburg_spec}{pinto_absolute}.
The fitted elemental compositions are in agreement with the expected
values in all cases.

\begin{figure}
\centering{}\subfloat[5.2 mm shifter on left half of field, see \figref{wsh52}.]{\includegraphics[width=0.49\columnwidth]{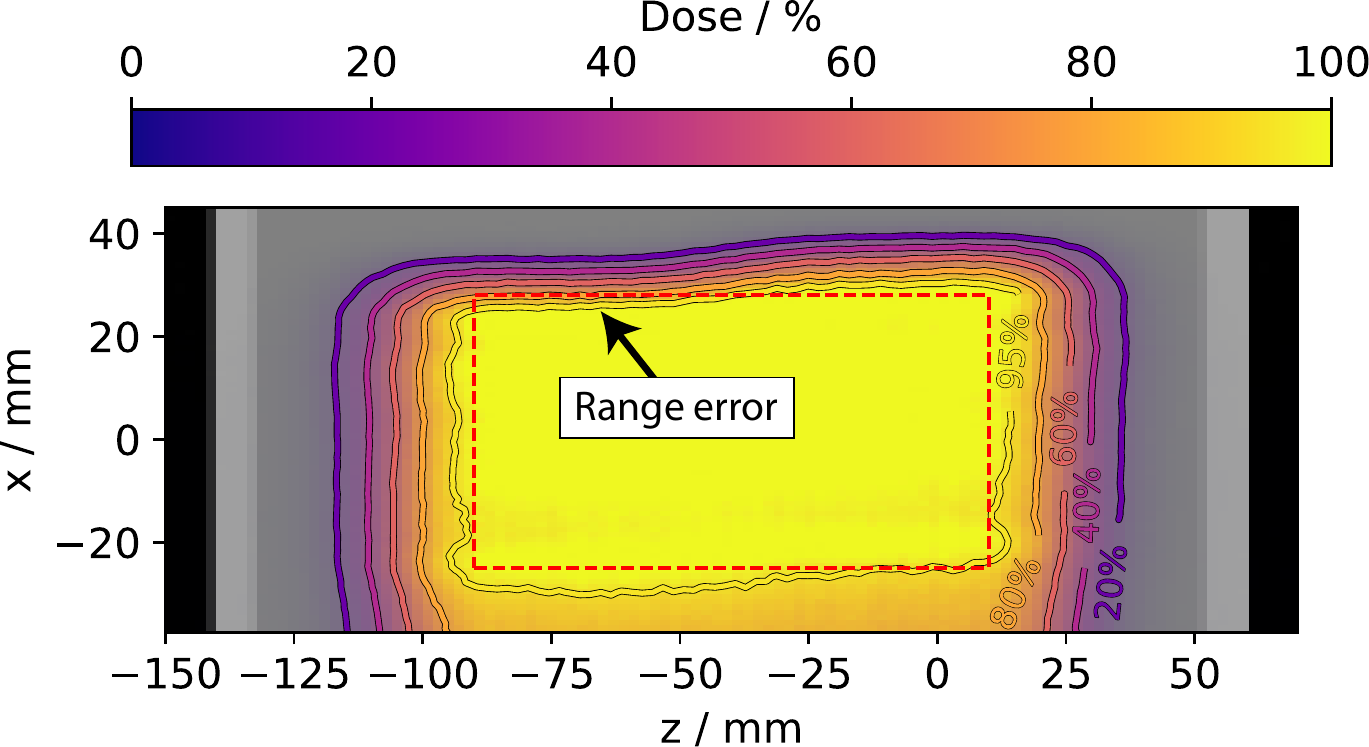}\label{fig:dosewsh52}}\subfloat[Nominal case, see \figref{nsh}.]{\includegraphics[width=0.49\columnwidth]{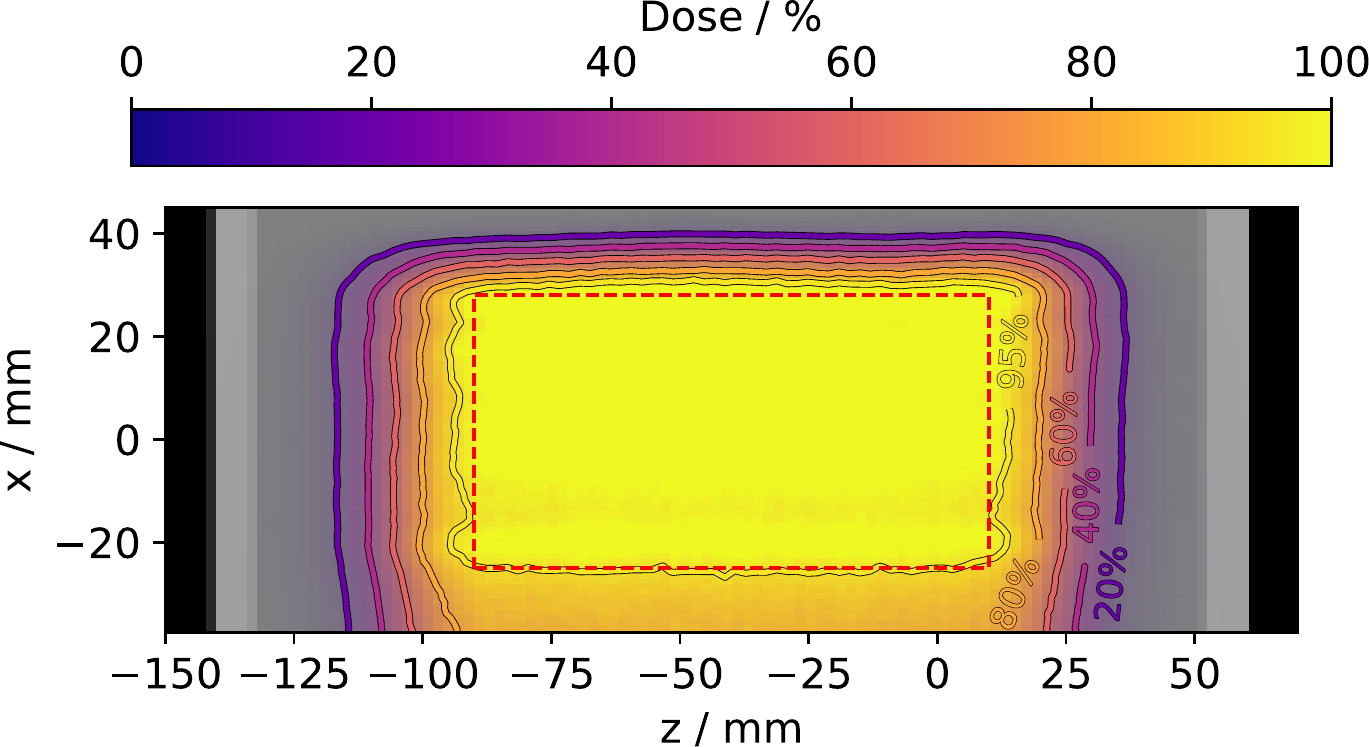}\label{fig:dosensh}}\textsf{\caption{Planning images superimposed with the reconstructed proton dose based
on the measurements of \figref{nsh-hsw-wsh-wbi}. The pencil-beams
are simulated with the GPU based on the measured proton range. Axes
are in mm and correspond to the FOV definition (\figref{fov}). The
red dashed rectangle shows the target region of 5.3 cm $\times$ 10
cm covered uniformly with 0.9 Gy. The image corresponds to slice $y=0$
mm. Isodose lines at 95\%, 80\%, 60\%, 40\% and 20\% levels of the
target dose are shown.}
\label{fig:doses} }
\end{figure}

We also recalculated the dose distributions based on the range verification
results. In \figref{doses}, the central slice $y=0$ mm of the dose
distribution in the water phantom is shown. The setup with a 5.2 mm
range shifter on the left part of the field (\figref{wsh52}) is compared
to the reference case (\figref{nsh}). The underdosage in the distal
part of the target volume (red dashed rectangle) and the overdosage
in the proximal region, that are introduced by the range shift, can
be observed.

\section{Discussion\label{sec:Discussion}}

Compared to our forerunner pre-clinical prototype with a single detector
\parencite{verburg_spec}, we have further developed and scaled up
our prompt gamma-ray spectroscopy technology to achieve millimeter
accuracy in proton range verification at a clinical beam current and
dose. Even under these challenging conditions, our system is uniquely
able to perform energy- and time-resolved prompt gamma-ray measurements
that are normalized to absolute units, which facilitates a direct
comparison with nuclear reaction models. The statistical precision
achieved with this prototype outperforms previously published results
from other prompt gamma-ray imaging systems at clinical beam currents
and doses, mainly because of the higher detector throughput and efficiency.

Prompt gamma-ray spectroscopy has the advantage of incorporating an
energy- and time-resolved event analysis in addition to the spatial
collimation. The range verification algorithm is therefore able to
reconstruct the proton range and the elemental composition of the
target tissue simultaneously. This reduces the bias in the retrieved
proton range by minimizing the uncertainty in the tissue composition,
which can be different from the estimation based on the planning CT
scan due to conversion ambiguity. A potential future improvement is
to incorporate dual-energy CT for characterization of the elemental
composition \parencite{wohlfahrt_dect}. This would improve the accuracy
of our model and it may be used to set constraints on the elemental
concentrations $\rho^{vt}$.

Our method considers many different nuclear reaction channels, which
yields redundant information for a reliable range verification. The
model enables the prediction of the absolute proton range with a bias
below 1 millimeter only based on the planning CT, the measured gamma-rays
and fundamental physical principles. The simulation is based on established
Monte Carlo methods, tabulated nuclear cross sections, and an empirical
calibration of the unresolved proton continuum. This direct physical
basis overcomes one of the main limitations of range verification
using positron emission tomography, where biological washout effects
need to be considered, which are difficult to model \parencite{knopf_pet}.

The performance of our system is enhanced by the design of the collimator,
which is set up in such a way that the FOV is focused on the distal
edge of the Bragg peak, where the emitted prompt gamma-rays have a
stronger correlation with the range of the beam than in the entrance
path. The slit opening is chosen as a compromise between systematic
uncertainties and statistical precision: a narrower collimator would
increase the spatial resolution but impair the collected statistics
in detector row $r=0$. The open slit configuration for the distal
detector row $r=1$, as previously shown by \cite{verburg_head},
maximizes the number of collected events at the end-of-range. In an
actual patient, where the beam entrance surface might not be flat
and the tissue is not homogeneous, the proton ranges of the spots
from the same energy layer will not be at the same depth with respect
to the collimator. Hence, having two detector rows at different depths
will be beneficial to enclose most of the proton stopping points within
the FOV.

The range errors that we introduced in this work were range undershoots.
In the case of a range overshoot, the statistical precision will be
higher because of more gamma-rays reaching the detectors. Also, as
expected from the prompt gamma-ray yields, the statistical precision
is higher in matter with a high \textsuperscript{16}O concentration
as compared to matter with a high \textsuperscript{12}C concentration.
This is advantageous for clinical range verification. The oxygen fraction
in a variety of tumor samples has been shown to range between 57\%
and 78\% \parencite{maughan_tumorcomposition}. Brain tissue and soft
tissue, which are often critical organs, are also rich in oxygen.

The aggregation of prompt gamma-ray measurements from proton pencil-beams
within a 10 mm lateral merging radius and a 10 mm depth is what we
believe constitutes a good trade-off between resolution and statistical
precision. The data that we used for the range verification of each
distal spot correspond to the delivery of about $1.6\times10^{9}$
protons and a mean accumulated dose to the target volume of 1.6 cGy.
In the case of single-field optimization of the dose delivery, as
was used in the experiments in this work, the large majority of the
protons stop very close the distal end of the target. Therefore, the
verification of the range of the distal proton energy layers is of
main importance. Multi-field optimized treatment plans created with
robust optimization also deliver most dose to the distal layers, because
the delivery of additional dose outside the target volume is possible
there to account for range uncertainty. When prompt gamma-ray spectroscopy
allows for a significant reduction of the range uncertainty, intensity-modulated
treatment plans may be designed with higher doses being delivered
to parts of the more proximal energy layers. In this situation, it
will likely be advantageous to use a spatially varying merging region
to account for the local differences in the delivered number of protons.

\begin{figure}
\begin{centering}
\includegraphics[width=0.8\columnwidth]{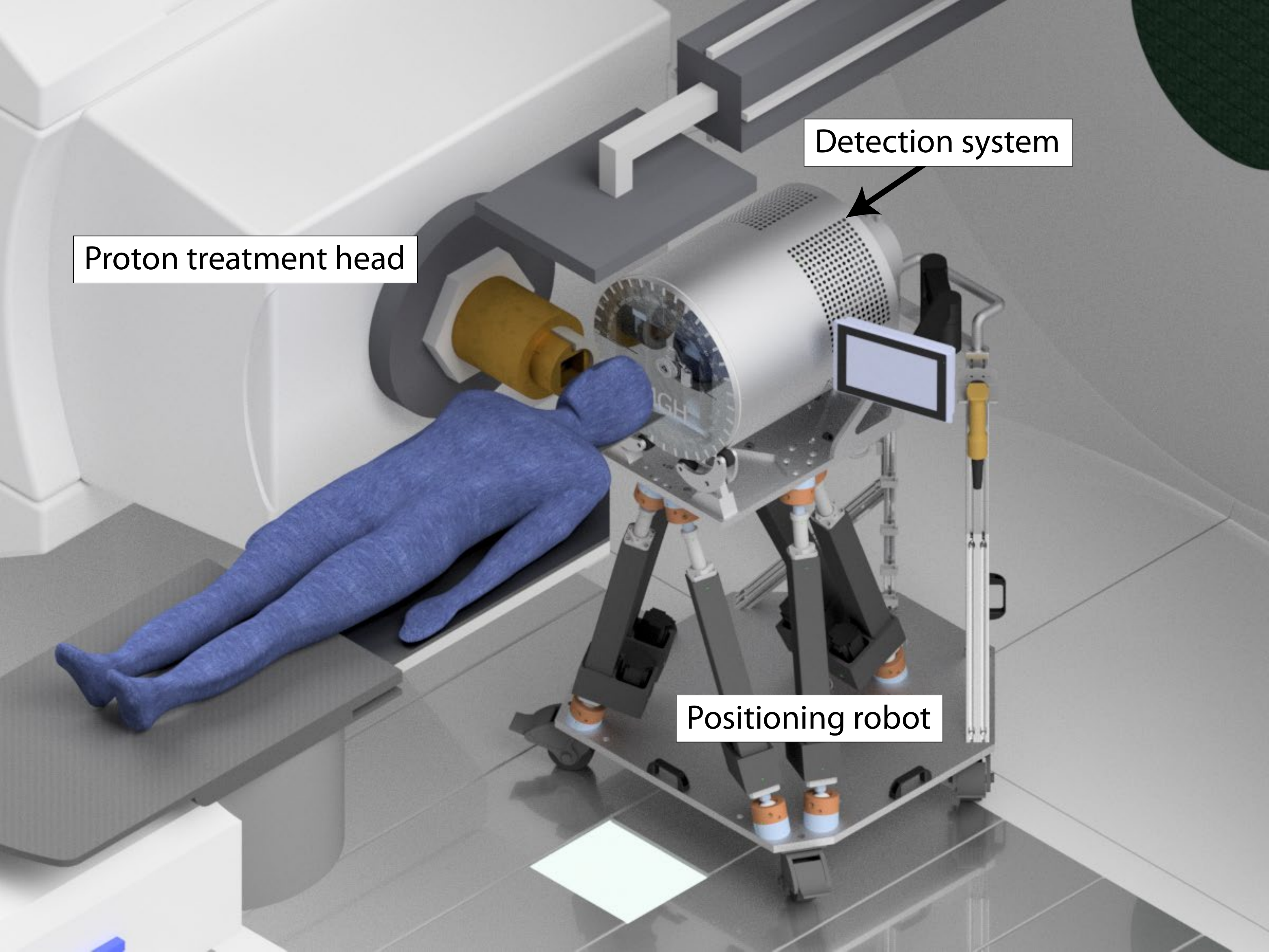}
\par\end{centering}
\textsf{\caption{Illustration of the gantry, proton treatment head, patient, top x-ray
flat panel and our prototype range verification system. The detector
frame can rotate according to the gantry angle, and is mounted on
a positioning robot, consisting of six actuators. The robot stands
on a platform on wheels that is moved into the treatment room.}
\label{fig:robot} }
\end{figure}

The prompt gamma-ray spectroscopy method is able to estimate the real
composition, as well as the magnitude of the range error, even if
the delivered pencil-beam ranges are quite different from the treatment
plan. Several potential range errors are simulated and the most likely
one according to the match with the measurements is determined. It
is however possible that complex differences occur between the treatment
planning and the delivery, for example, if the heterogeneities in
the beam path are very different as compared to the plan. In this
case, significant discrepancies between the model and the measurement
data would be observable, which might be more difficult to interpret
as only a range error. This would serve as an indication that the
positioning of the patient has to be checked and that re-planning
may be necessary. When adaptive workflows based on a daily cone-beam
CT become more commonplace, this would be less of an issue.

For a smooth integration of our prototype in the clinical workflow
during the first patient study, we are developing a six-axis positioning
robot and a laser alignment system. Our rotating detector frame will
be mounted on these actuators as shown in \figref{robot}, which will
enable a fast, accurate and reproducible detector alignment with respect
to the patient using several alignment lasers. The robot itself will
stand on a mobile platform.

The reconstructed proton range and dose will be superimposed on the
patient CT to assess for significant range deviations with respect
to the treatment plan, as shown in \figref{dosewsh52}. As emphasized
by \textcite{verburg_spec}, the absolute range can be determined
without prior knowledge of the tissue composition already for the
first treatment fraction, and further relative range shifts can be
detected for later fractions with high precision.

Once the robustness and accuracy of the prototype is demonstrated
during actual treatments, the prompt gamma-ray spectroscopy system
could promote the correction of range errors in upcoming treatment
fractions with re-planning, thus improving the quality of treatment
and, potentially, the outcome. On a longer term, we envision real-time
adaptation of the range, the reduction of range margins, and novel
treatment plan designs that take advantage of the sharp distal gradient
of the Bragg peak to maximize the sparing of organs-at-risks that
surround the target volume.

\section{Conclusions\label{sec:Conclusions}}

A full-scale clinical prototype system for proton range verification
using prompt gamma-ray spectroscopy was tested with phantoms at the
pencil-beam scanning gantry treatment room of the Francis H. Burr
Proton Therapy Center. The developed electronics and calibration methods
performed well at a clinical beam current of 2 nA incident on the
phantoms. A detailed model of prompt gamma-ray emission and system
response was developed, that enables the reconstruction of both the
proton range and the elemental composition of the irradiated matter
based on the measurements.

The absolute range was determined for each pencil-beam spot of the
distal energy layer with a mean statistical precision of 1.1 mm at
a 95\% confidence level and a mean absolute deviation of 0.5 mm, for
a delivered dose of 0.9 Gy, and by aggregating data from pencil-beams
in a merging cylinder of 10 mm radius and 10 mm depth. Range errors
that we introduced were detected accurately even in the presence of
large differences in the elemental composition with respect to the
assumptions based on the planning CT scan.

Our results show for the first time that proton range verification
with 1 millimeter precision is achievable in phantoms under clinically
realistic conditions. An experiment with a ground-truth anthropomorphic
head phantom \parencite{wohlfahrt_burns} is planned as the final
validation prior to our upcoming clinical study, where we will test
our prototype during the treatment of patients with brain tumors.

\ack{}{}

We thank Arthur Brown, Pablo Botas, Ethan Cascio, Bob Brett, Tom Madden,
Georgios Patoulidis and the staff of the proton therapy facility for
their excellent support. This work was supported in part by the Federal
Share of program income earned by Massachusetts General Hospital on
C06-CA059267, Proton Therapy Research and Treatment Center, and by
the National Cancer Institute grant U19 CA021239-35. The Tesla K40
GPU was donated by the NVIDIA Corporation.

\printbibliography[heading=bibintoc]

\end{document}